 \documentclass[3p]{elsart}

\usepackage{multirow}
 \usepackage{epstopdf}
\usepackage{amsmath,url,subfigure}

\usepackage{lineno}

\usepackage[usenames,dvipsnames]{color}
\usepackage{hyperref}
\hypersetup{colorlinks=true}
\hypersetup{linkcolor=DarkOrchid}
\hypersetup{citecolor=DarkOrchid}        
\hypersetup{urlcolor=DarkOrchid}

\begin{document}
\begin{frontmatter}
\title{Energy and Flux Measurements of Ultra-High Energy Cosmic Rays Observed During the First ANITA Flight}
\author[UH,MPIK]{H.~Schoorlemmer\corref{cor1}}
\ead{harmscho@mpi-hd.mpg.de}
\cortext[cor1]{Corresponding author}

\author[UCLA,JPL]{K.~Belov}
\author[JPL]{A.~Romero-Wolf}
\author[USC]{D. Garc{\'\i}a-Fern\'andez}
\author[Wash]{V.~Bugaev}
\author[UCLA,CalPal]{S.~A.~Wissel}
\author[OHIO]{P.~Allison}
\author[USC]{J. Alvarez-Mu\~niz}
\author[Irvine]{S.~W.~Barwick}
\author[OHIO]{J.~J.~Beatty}
\author[Kansas,Moscow]{D.~Z.~Besson}
\author[Wash]{W.~R.~Binns}
\author[USC]{W.~R.~Carvalho Jr.}
\author[Tai]{C.~Chen}
\author[Tai,SLAC]{P.~Chen}
\author[Delw]{J.~M.~Clem}
\author[OHIO]{A.~Connolly}
\author[Wash]{P.~F.~Dowkontt}
\author[UH]{M.~A.~DuVernois}
\author[SLAC]{R.~C.~Field}
\author[Irvine]{D.~Goldstein}
\author[UH]{P.~W.~Gorham}
\author[SLAC]{C.~Hast}
\author[KIT]{T.~Huege}
\author[UH]{C.~L.~Heber}
\author[UCLA]{S.~Hoover}
\author[Wash]{M.~H.~Israel}
\author[Delw]{A.~Javaid}
\author[UH]{J.~Kowalski}
\author[UCLA]{J.~Lam}
\author[UH]{J.~G.~Learned}
\author[Wash]{J.~T.~Link}
\author[UH]{E.~Lusczek}
\author[UH]{S.~Matsuno}
\author[OHIO]{B.~C.~Mercurio}
\author[UH]{C.~Miki}
\author[UH]{P.~Mio\v{c}inovi\'c}
\author[Delw]{K.~Mulrey}  
\author[UH]{J.~Nam}
\author[JPL]{C.~J.~Naudet}
\author[SLAC]{J.~Ng}
\author[UCL]{R.~J.~Nichol}
\author[OHIO]{K.~Palladino}
\author[Wash]{B.~F.~Rauch}
\author[UH]{J.~Roberts}
\author[SLAC]{K.~Reil}
\author[UH]{B. Rotter}
\author[UH]{M.~Rosen}
\author[UH]{L.~Ruckman}
\author[UCLA]{D.~Saltzberg}
\author[Delw]{D.~Seckel}
\author[UCLA]{D.~Urdaneta}
\author[UH]{G.~S.~Varner}
\author[Chicago]{A.~G.~Vieregg}
\author[SLAC]{D.~Walz}
\author[Irvine]{F.~Wu}
\author[USC]{E.~Zas}
\vspace{2mm}
\noindent

\address[UH]{University of Hawaii at Manoa, 
Department of Physics and Astronomy, Honolulu,
Hawaii 96822, USA.}
\address[MPIK]{Max-Planck-Institut f\"ur Kernphysik, 69117, Heidelberg, Germany} 
\address[UCLA]{Dept. of Physics and Astronomy, Univ. of California, Los Angeles, Los Angeles CA
90095, USA}
\address[JPL]{Jet Propulsion Laboratory, California Institute of Technology, Pasadena, CA 91109. USA}
\address[USC]{Departamento de F\'\i sica de Part\'\i culas
\& Instituto Galego de F\'\i sica de Altas Enerx\'\i as,
Universidade de Santiago de Compostela,\\ 
15782 Santiago de Compostela, Spain.}
\address[Wash]{Dept. of Physics and McDonnell Center for the Space Sciences, Washington Univ. in St. Louis, MO 63130,USA}
\address[CalPal]{Dept. of Physics, California Polytechnic State Univ., San Luis Obispo CA 93407, USA}
\address[OHIO]{Dept. of Physics, Ohio State Univ., Columbus, OH 43210, USA}
\address[Irvine]{Dept. of Physics, Univ. of California, Irvine, CA 92697, USA}
\address[Kansas]{Dept. of Physics and Astronomy, Univ. of Kansas, Lawrence, KS 66045, USA }
\address[Moscow]{National Research Nuclear University (Moscow Engineering Physics Institute),
Moscow Russia, 115409}
\address[Tai]{Dept. of Physics, Grad. Inst. of Astrophys.,\& Leung Center for Cosmology and Particle Astrophysics, National Taiwan University, Taipei, Taiwan.}
\address[Delw]{Dept. of Physics, Univ. of Delaware, Newark, DE 19716, USA }
\address[KIT]{Institut f\"ur Kernphysik, Karlsruhe Institute of Technology (KIT), Germany}
\address[UCL]{Dept. of Physics and Astronomy, University College London, London, United Kingdom. }
\address[SLAC]{SLAC National Accelerator Laboratory, Menlo Park, CA, 94025, USA }
\address[Min]{School of Physics and Astronomy, Univ. of Minnesota, Minneapolis, MN 55455, USA.}
\address[Chicago]{Dept. of Physics, Enrico Fermi Institute, Kavli Institute for Cosmological Physics, Univ. of Chicago , Chicago IL 60637, USA}

\begin{abstract}
The first flight of the Antarctic Impulsive Transient Antenna (ANITA) experiment recorded 16 radio signals that were emitted by cosmic-ray induced air showers. The dominant contribution to the radiation comes from the deflection of positrons and electrons in the geomagnetic field, which is beamed in the direction of motion of the air shower. For 14 of these events, this radiation is reflected from the ice and subsequently detected by the ANITA experiment at a flight altitude of $\sim$36\,km. In this paper, we estimate the energy of the 14 individual events and find that the mean energy of the cosmic-ray sample is $2.9 \times10^{18}$\,eV, which is significantly lower than the previous estimate. By simulating the ANITA flight, we calculate its exposure for ultra-high energy cosmic rays. We estimate for the first time the cosmic-ray flux derived only from radio observations and find agreement with measurements performed at other observatories. In addition, we find that the ANITA data set is consistent with Monte-Carlo simulations for the total number of observed events and with the properties of those events.
\end{abstract}

\end{frontmatter}

\section{Introduction}
Ultra-high energy cosmic rays (UHECRs) have been observed for over half a century; however, their nature and origin remain uncertain. Modern observatories, like the Pierre Auger Observatory in Argentina \cite{ref:AugerSD,ref:AugerFD} and the Telescope Array in Utah \cite{ref:TASD,ref:TAFD}, try to unravel the mysteries of UHECRs not only by increasing the number of recorded UHECRs, but also by increasing the measurement accuracy of their properties. Simultaneous observations of the fluorescence light emitted by the air shower while it passes through the atmosphere and the particle footprint on the ground provide cross calibration of the primary particle's properties including its energy. This method significantly reduces the systematic uncertainties in the observation of UHECRs. However, there still remains a discrepancy between the observed cosmic-ray energy spectrum at the Pierre Auger Observatory and the Telescope Array (see for example \cite{ref:TA}). In this paper, we develop an independent method to estimate the energy of UHECRs and apply this to cosmic rays observed during the first Antarctic Impulsive Transient Antenna (ANITA) flight, ANITA-I.\par
When the cosmic-ray sample of the first ANITA flight was reported \cite{ANITA_CR} none of the models for calculating radio emission from air showers was able to predict significant radiation in the frequency band of ANITA-I. A preliminary energy estimate of this sample was performed using a Bayesian approach with a simple model as the prior for the emission pattern. In this paper the cosmic-ray data set is re-analyzed using a realistic model for the radio emission of air showers in order to estimate the cosmic-ray energy. The analysis presented here shows consistency with cosmic-ray flux observations made with other experiments and results in consistency when comparing distributions of parameters obtained from a full flight simulation to the observed distributions. The resulting energy scale of the cosmic-ray sample is about a factor of four lower than the previous estimate.\par
We first introduce the ANITA-I experiment and briefly discuss its cosmic-ray data set. In the following section, we introduce our energy estimation method and apply it to the cosmic-ray observations. Thereafter, we perform a full Monte Carlo simulation of the ANITA-I flight to calculate the exposure and cosmic-ray flux and test for consistency between predictions from simulations and cosmic-ray observations. We conclude with a discussion of the obtained results and a brief discussion as to how these results could impact future experiments.

\section{The cosmic-ray data set obtained with ANITA-I}
The objective of the ANITA-I experiment was to observe impulsive Askaryan radiation resulting from ultra-high-energy neutrino interactions in the Antarctic ice. However, during its first flight the ANITA experiment collected a data set of 16 radio signals that were emitted by cosmic-ray air showers \cite{ANITA_CR}. These events remained after careful rejection of man-made pulsed signals, continuous waves, and thermal noise events. They all exhibited the polarization signature typical of the geomagnetic radiation from cosmic-ray air showers \cite{ref:CodGeo,ref:AugePol}.\par 
The arrival directions of two events were from above the horizon. This indicated that these signals came directly from air showers that skimmed through the atmosphere without reaching the Earth's surface. The arrival directions of the radiation of the other 14 events pointed back toward locations on the Antarctic ice sheet. The polarity of the signal was inverted with respect to the two direct events. These signals came from down-going air showers where the radio emission is reflected off the Antarctic ice sheet toward the payload.  Understanding the particle shower and the resulting radiation from the two direct events is very interesting. These direct events correspond to particle showers which develop in significantly lower atmospheric density and never reach the ground. Therefore, the shower development is expected to differ significantly from standard down-going particle showers. To understand these particle showers, and the resulting radiation from them, dedicated shower simulations are required, which means modification of existing simulation codes. This is out of the scope of the analysis  presented in this paper. 

\section{Description of the ANITA instrument and its first flight}
The ANITA-I experiment was suspended from a high altitude balloon that flew in the 2006-2007 Austral summer over Antarctica. The total flight took 35 days in which the payload made almost four revolutions around the south pole. However, due to stability issues with the flight computer and periods in high background environments (near the base after take off) the effective uptime of the experiment was 17.25 days. The payload had a float altitude of $\sim$36\,km and an antenna array provided a panoramic view on the ice sheet below it. The antenna array consisted of 32 dual-polarized quad-ridged horn antennas optimized to observe radiation in the 200-1200\,MHz band. Each antenna had a vertical and horizontal polarization feed, and the full-width-half-maximum beamwidth was about 45$^{\circ}$. The antennas were arranged in an azimuthally-symmetric array with one ring of 16 antennas at lower part of the payload and two rings of antennas  (8+8) at the top of the payload. Each antenna is tilted 10$^{\circ}$ downward.\par
 Signals from each antenna feed were amplified and bandpass-filtered into 64 channels. These channels were then split into a trigger channel and a channel for digitization. The signal going into the trigger channel was decomposed into left- and a right-handed circular polarization. The two circularly-polarized signals were split into four frequency bands that were connected to tunnel diodes that functioned as power detectors. A trigger signal was issued when a threshold crossing was observed in several frequency bands (three or more out of eight). When a combination of neighboring antennas (minimal 4) in the array issued a trigger in time coincidence, signals from all of the antennas in the array were digitized and written to disk for a duration of 100\,ns  and with a sampling frequency of 2.6\,GSa\,s$^{-1}$.\par
Orientation, timing, and location information was provided by an array of GPS units. In addition, redundant systems using sun sensors and magnetometers were used to provide additional orientation information. Trigger efficiency and direction reconstruction were validated inflight by ground pulser stations. For a more detailed description of the instrument and its performance see \cite{ref:AnitaExp}.\par

\section{Energy estimation of individual events}
\subsection{Parameters obtained from the observed signals}
A plane wave arriving at the antenna array will induce a pattern of arrival times corresponding to the incoming direction. We use these relative arrival times to calculate the coherent power corresponding to a certain direction. By calculating the coherent power from multiple directions a two dimensional interferometric map is generated. The location of the maximum coherent power within this map is used as the reconstructed incoming direction of the signal. This technique is explained in more detail in \cite{ref:AndresInfer}.\par
\begin{figure}
\begin{center}
{
\includegraphics[width=0.45\textwidth]{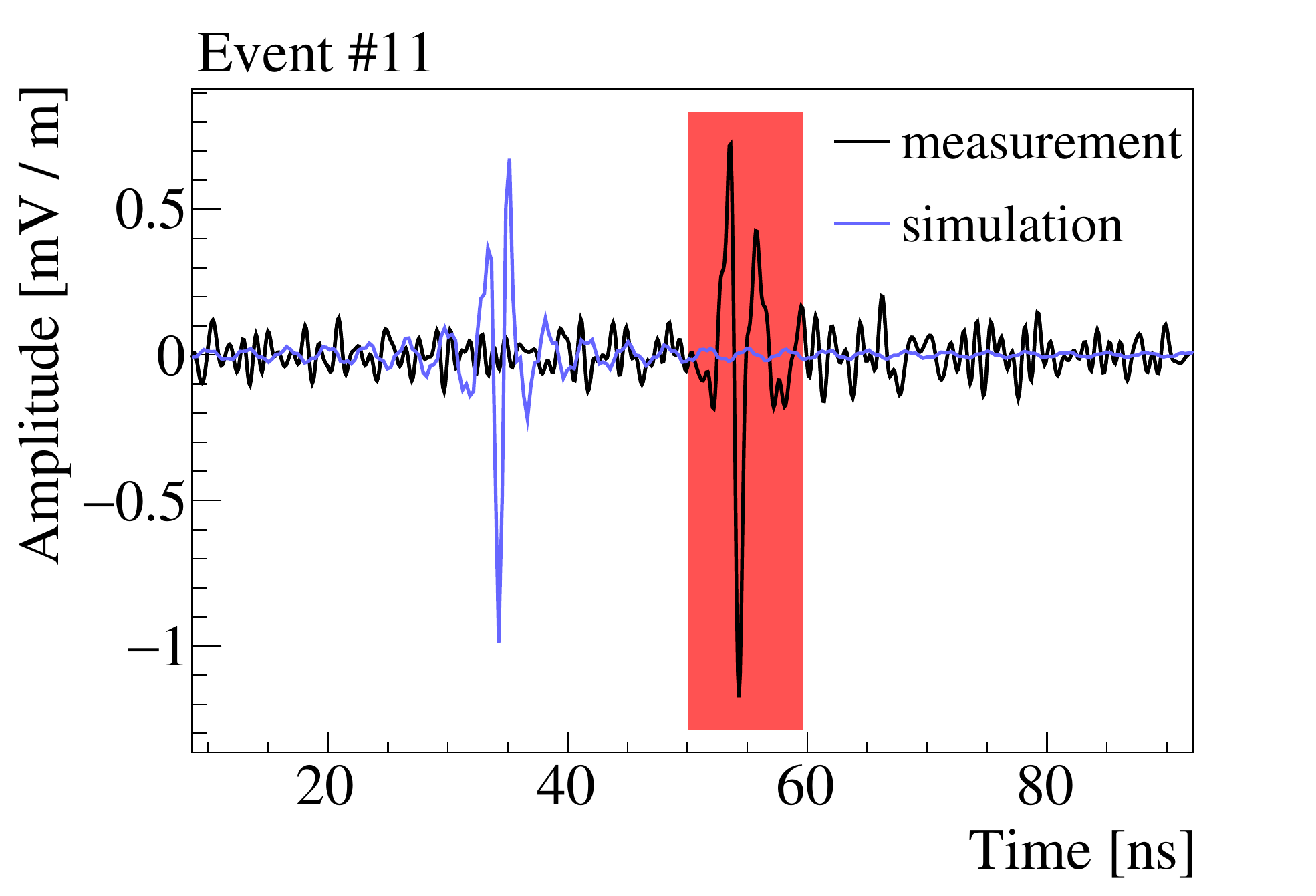}
\includegraphics[width=0.45\textwidth]{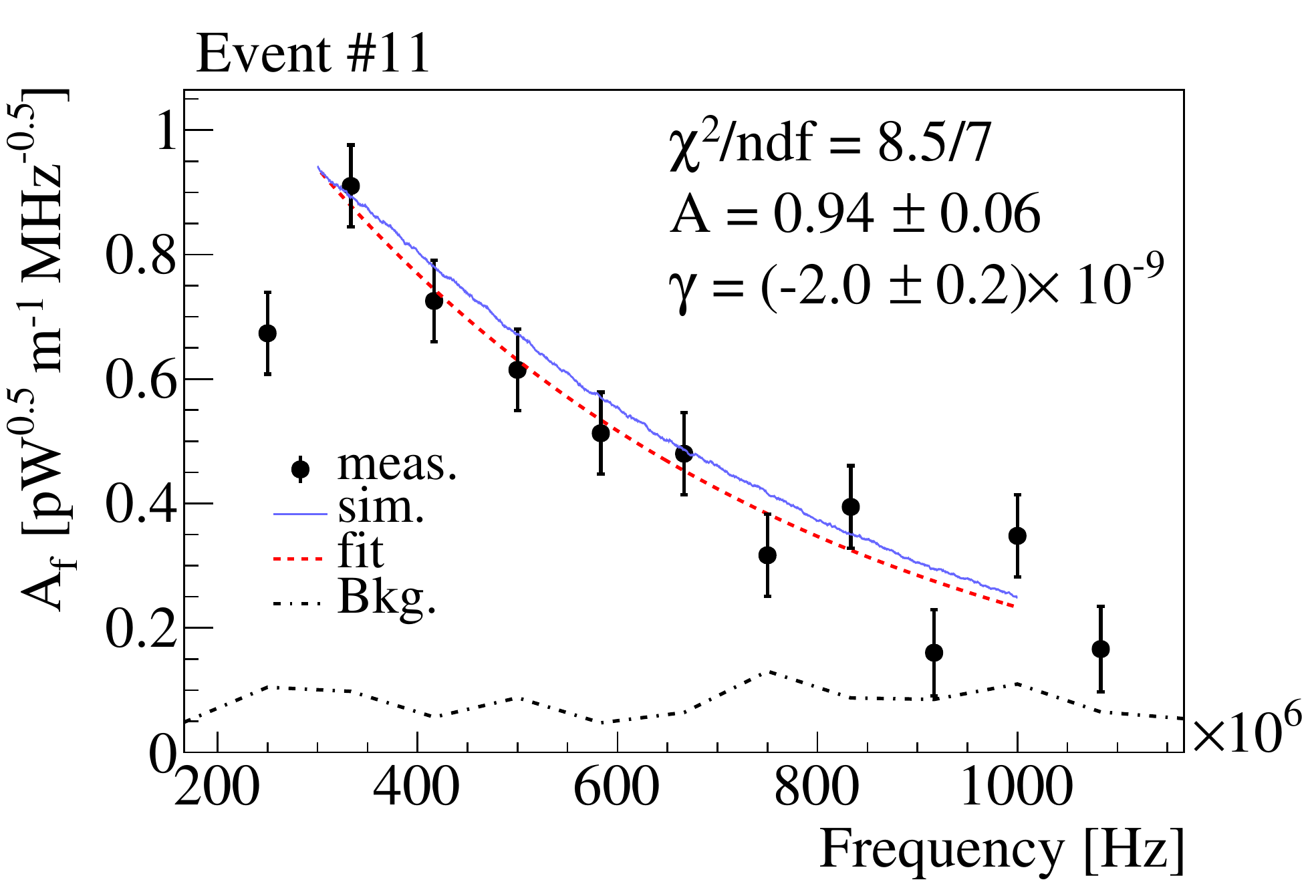}
}
\caption{Left: The time dependent electric field induced by an air shower at the location of the ANITA payload. The impulsive signal is contained within a window of about 10\,ns as indicated in red. Right: The background corrected amplitude spectrum of the signal window. The black markers represent the estimated signal and their 68\,\% confidence interval (see text for details). A simple exponential fit (see Eq. (\ref{eq:exp})), in the range 300-1000\,MHz, to the signal spectrum is also shown. In addition we show an example of simulated radio pulse obtained with the ZHAireS Monte Carlo (with an arbitrary time-offset).}
\label{fig:ExampleSignal}
\end{center}
\end{figure}

After direction reconstruction, the electric field at the antenna feed is estimated by deconvolving the response function of both the signal chain and the effective height of the antennas from the measured voltage. Note that the effective height of each antenna depends on both frequency and the incoming direction of the radiation.
As an example of the reconstructed time-dependent electric field at the payload we show the calibrated measurement of one cosmic-ray signal in the left panel of Figure \ref{fig:ExampleSignal}.
 As indicated, the pulsed signal is fully contained within a window of about 10\,ns. The samples in this window are used to obtain the amplitude spectrum as shown in the right panel of Figure \ref{fig:ExampleSignal}. \par
The amplitude spectrum below 300\,MHz suffers from loss of power due to filtering of emission from satellites between 220 and 290\,MHz. In addition, the antenna model is less accurate at the lower frequencies. The antenna model is obtained from preflight measurements, and the lab setup provides less accurate results for the low frequency part of the antenna sensitivity due to near field effects. Therefore, frequencies below 300 are not considered in the spectral fitting.  Around 1000\,MHz, the sensitivity of the digitizer drops rapidly, which makes signal recovery above this frequency unreliable. Therefore, we choose to evaluate the amplitude spectrum between 300 and 1000\,MHz.\par 
The observed amplitude at a given frequency is the sum of the signal and a thermal noise background with a random phase. The probability distribution of the resulting amplitude is described by a Rician distribution. We use this probability distribution function to estimate a central 68\,\% confidence interval in which the signal is contained given the measured amplitude and the background. The background is estimated from several (4 or 5) 10\,ns time windows in the same waveform as the signal while excluding the signal region. We average the amplitude spectrum from these sideband samples to get a frequency-dependent amplitude spectrum of the background. We observed that within statistical fluctuation this background amplitude spectrum is flat over the band (as shown in Figure \ref{fig:ExampleSignal} (right)). We then averaged the background over the full amplitude spectrum to get a single estimation of the background level for the complete frequency band. The range of observed background levels are between 0.07 and 0.2\,pW$^{0.5}$m$^{-1}$MHz$^{-0.5}$.\par
From the signal spectrum, $A_f$, as a function of frequency, $f$, we obtain two parameters by fitting a simple exponential function
\begin{equation}
A_f(f) = A e^{\gamma(f - 300\text{\,MHz})}.
\label{eq:exp}
\end{equation}
Parameter $A$ gives the amplitude of the electric field at 300\,MHz (at the location of ANITA) and parameter $\gamma$ describes the frequency dependence of the amplitude spectrum. An exponential function is motivated by the shape of the amplitude spectrum obtained from ZHAireS simulations \cite{ref:ZHAireSCher}. 
The fit parameters for the 14 cosmic-ray observations are listed in Table \ref{tab:FitParameters}.

\begin{table}[htdp]
\begin{center}
\begin{tabular}{|c|c|c|c|}
\hline
Event & $\theta$ 		&$A$ 	 		&$\gamma$  \\
numbers	 	& $[^{\circ}]$ 	&$[\rm pW^{0.5}~m^{-1}~MHz^{-0.5}]$ &[Hz$^{-1}$] \\
\hline
1				&  	84.6				&0.25	$\pm$ 0.03	 			&-1.6 $\pm	~0.5\times 10^{-9}$ \\
2				& 	80.4				&0.72	$\pm$ 0.04	 			&-2.2 $\pm~	0.2 \times 10^{-9}$ \\
3				&		65.5				&0.75	$\pm$ 0.08 				&-2.3 $\pm~	0.5 \times 10^{-9}$ \\ 
4				&		65.6				&0.71	$\pm$ 0.06				&-2.5 $\pm~0.4 \times  10^{-9}$ \\
5				&		64.0				&2.90	$\pm$ 0.13				&-3.1 $\pm	~0.2\times 10^{-9}$ \\
6				&		68.7				&0.73 $\pm$ 0.04 				&-1.7 $\pm	~0.2\times 10^{-9}$ \\
7				&		74.9				&0.31 $\pm$ 0.04	 			&-2.0 $\pm	~0.6\times 10^{-9}$ \\
8				&		57.0				&0.66 $\pm$ 0.13 				&-3.2 $\pm	~1.5\times 10^{-9}$ \\
9				&		74.5				&0.34 $\pm$ 0.05 				&-2.3 $\pm	~0.6\times 10^{-9}$ \\ 
10			&		78.8				&0.61 $\pm$ 0.05				&-3.2 $\pm	~0.4\times 10^{-9}$\\
11			&		70.5				&0.94 $\pm$ 0.06				&-2.0 $\pm	~0.2\times 10^{-9}$\\
12			&		79.1				&0.39 $\pm$ 0.05				&-2.4 $\pm	~0.6\times 10^{-9}$\\
13			&		81.9				&0.72 $\pm$ 0.06				&-4.0 $\pm~0.6\times 10^{-9}$\\
14			&		78.6			  	&0.57 $\pm$ 0.04				&-2.2 $\pm~0.3\times 10^{-9}$ \\\hline
\end{tabular} 
\caption{Incident angle of the radiation on the ice, $\theta$, and the fitted amplitudes at 300\,MHz ($A$) and spectral slopes ($\gamma$) (see Eq. (\ref{eq:exp})) for the 14 cosmic-ray events.}
\label{tab:FitParameters}
\end{center}
\end{table}
In addition to the measured pulse, we show in Figure \ref{fig:ExampleSignal} also an example of a ZHAireS simulation with an energy and off-axis angle close to reconstructed values (see section 4). A simple rectangular bandpass filter between 300 and 1000\,MHz is applied to the time domain of the simulations . The amplitude, polarity and duration of the simulated pulse are comparable to the measurement. However, small deviations of the pulse shape in the time domain are expected due to the simple bandpass filter applied to the simulation.

\subsection{Models for radio emission from air showers }
\begin{figure}
\begin{center}
{
\includegraphics[width=0.7\textwidth]{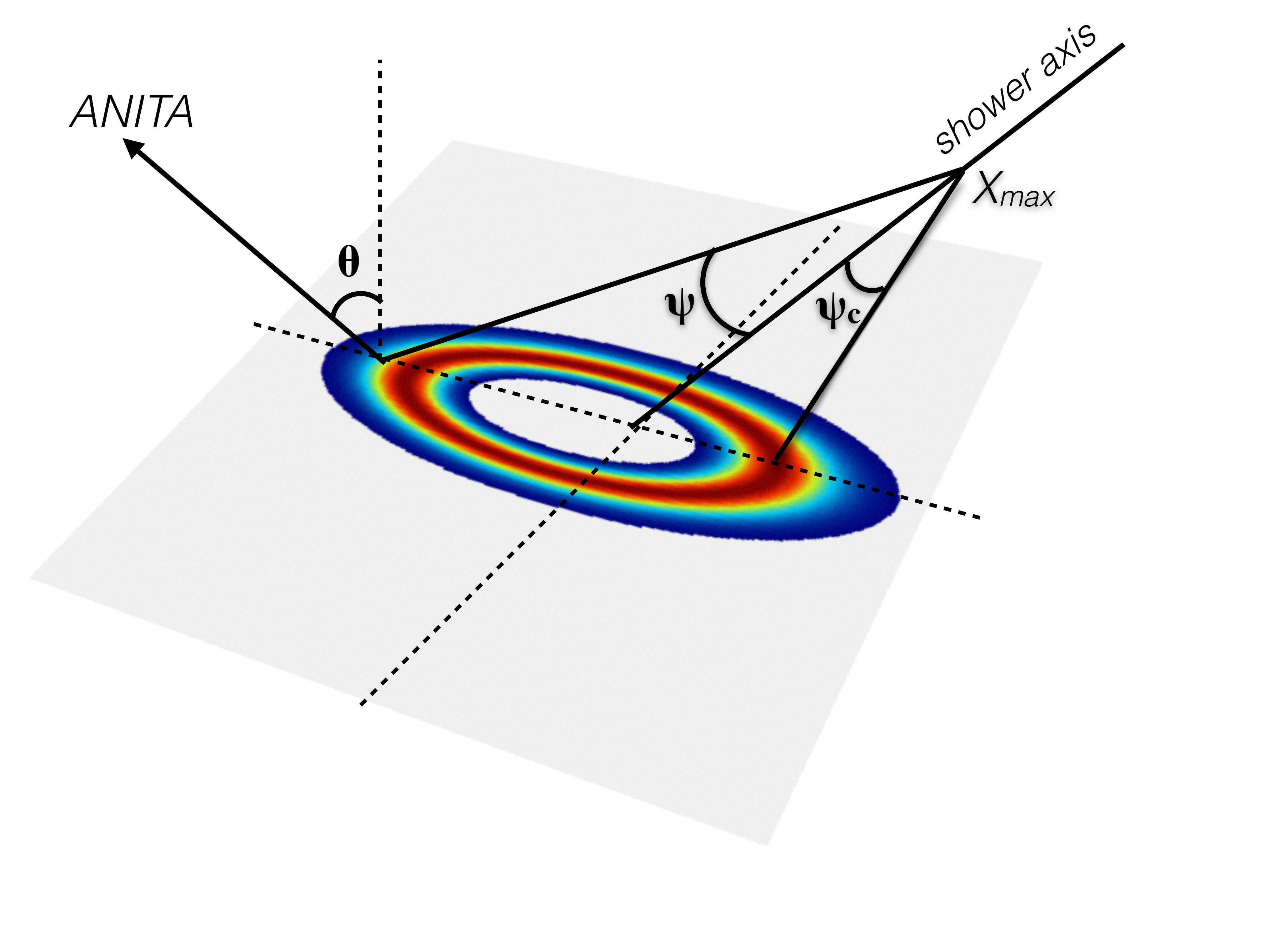}

}
\caption{Illustration of the footprint of the Cherenkov-ring on the ground and the geometry system used in the analysis. From the observation of radiation at the location of ANITA we can calculate the incident angle of the radiation on the ice $\theta$. After simulating an air shower, we can define a line from the location of $X_{\text{max}}$ that after reflection with angle $\theta$ would end up at the location of the payload. The angle this line makes with respect to the shower axis is defined as the off-axis angle $\psi$. A range of off-axis angles is probed for a single air shower simulation by varying the location of ANITA in simulation while keeping the payload at a fixed altitude above sea-level.  The off-axis angle of the Cherenkov ring is indicated by $\psi_c$.  The color map is obtained by calculating the radio signal at 300 MHz for an incoming cosmic ray with a zenith angle of $70^{\circ}$ using the ZHAireS code and the interpolation technique as discussed in \cite{ref:ZHS2com}. Note, due to the shape of the radiation pattern there will aways be an offset between the direction of the observed radiation and the direction of the shower-axis. This will be discussed in more detail in Section \ref{sec:corrections}.}
\label{fig:Geometry}
\end{center}
\end{figure}

During the ANITA-I flight in 2006-2007 none of the available models were able to predict observable signals in the 300-1000\,MHz frequency band. However, the recent inclusion of the effect of the atmospheric refractive index increased the frequency at which the simulations are coherent up to several gigahertz \cite{ref:ZHAireSCher,ref:KrijnCher,ref:CoREAS,ref:SELFAS}. This makes it now possible to compare the observation directly to the models and this is the main motivation to reanalyze the ANITA-I data set. The effect of the atmospheric refractive index on propagation of the radiation from the air shower results in a cone-like beam around the shower axis. The variation in arrival time of the radiation originating from different regions of the air shower shrinks to a minimum for observers located at the Cherenkov angle corresponding to the location of the region of maximum emission: Therefore, radiation with short wavelengths adds more coherently in the direction of the Cherenkov angle $\psi_c$, resulting in the flattest frequency spectrum at the Cherenkov cone. As an observer moves away from the Cherenkov angle, coherence is lost at the higher frequencies causing the frequency spectra to steepen. This frequency dependence as a function of distance to the Cherenkov angle plays a fundamental role in the method that we developed to estimate the energy of cosmic-ray particles. In the frequency band of the ANITA experiment the coherence due the Cherenkov effect, and therefore the atmospheric refractive index, is crucial to obtain measurable radiation. The dependence of the refractive index with altitude has an impact on the observed radiation. The refractive index falls off as a function of altitude. As a consequence, more inclined showers will emit their radiation within a smaller off-axis angle since they develop at higher altitude. 
At lower frequencies the conditions for coherent radiation are easier satisfied. Therefore, the time compression due to the Cherenkov-effect  does not play a dominant role anymore. As a result, the characteristic ring pattern is less prominent at lower frequencies. For examples of the shape of the radiation pattern over a large frequency range see \cite{ref:ZHSreflex}. The geometry for reflected radiation from a cosmic-ray-induced air shower is illustrated in Figure \ref{fig:Geometry}.\par
All of the standard simulation packages of radio emission from air showers have been developed to calculate the emission for antenna arrays on the ground. However, recently the ZHAireS code \cite{ref:ZHAireS} was upgraded \cite{ref:ZHSreflex} to incorporate a reflection off a dielectric medium, such as the Antarctic ice. The code reflects the radiation contribution from each particle track individually on the ice and propagates it to the payload location where they are summed to calculate the total electric field. Therefore, we can compare the simulated electric field properties directly to the observed ones at the location of the payload. At the reflection point the Fresnel coefficients are applied.\par 
This method avoids the problem of simulating the electric field in the near field on the ground, while the payload observes it in the far field. When both are in the far field one can simply extrapolate ground calculations to payload locations but this method does not apply in the range of incident angles as observed by ANITA (see \cite{ref:ZHSreflex}). This is why we had to use the upgraded ZHAireS package to properly calculate electric field strength at the payload. However, to study the dependency on the choice of simulation package we compared the ZHAireS simulation to another simulation package in Appendix B for antennas located on the ground.\par 

\subsection{Method for energy estimation}

In this section, we will explain the basic concept for estimating the cosmic-ray energy and then introduce the refinements that go into the final energy estimation. The method presented here builds upon concepts that were presented in \cite{ref:Kon,ref:ZAS}.\par
For a given simulated air shower we calculate the electric field at different off-axis angles, $\psi$ (see Figure \ref{fig:Geometry}). The falloff of the amplitude spectrum depends strongly on the off-axis angle and is reasonably described by a simple exponential in the frequency range of 300-1000\,MHz \cite{ref:ZHAireSCher}. As with the measurements, we fit the function in Equation (\ref{eq:exp}) to estimate the amplitude  $A$ at 300\,MHz and the spectral slope $\gamma$ at each off-axis angle.
\begin{figure}
\begin{center}
{
\includegraphics[width=0.45\textwidth]{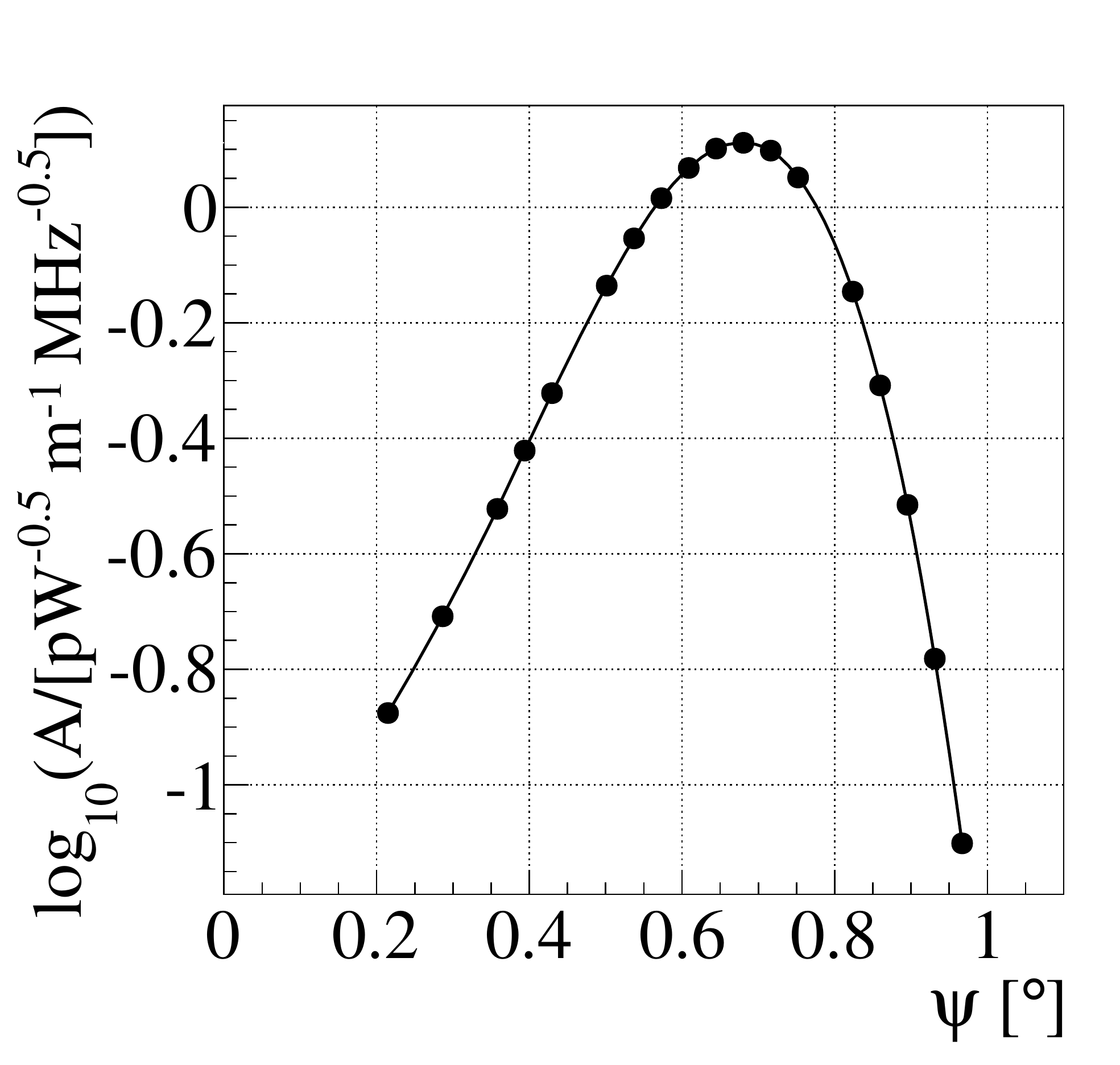} 
\includegraphics[width=0.45\textwidth]{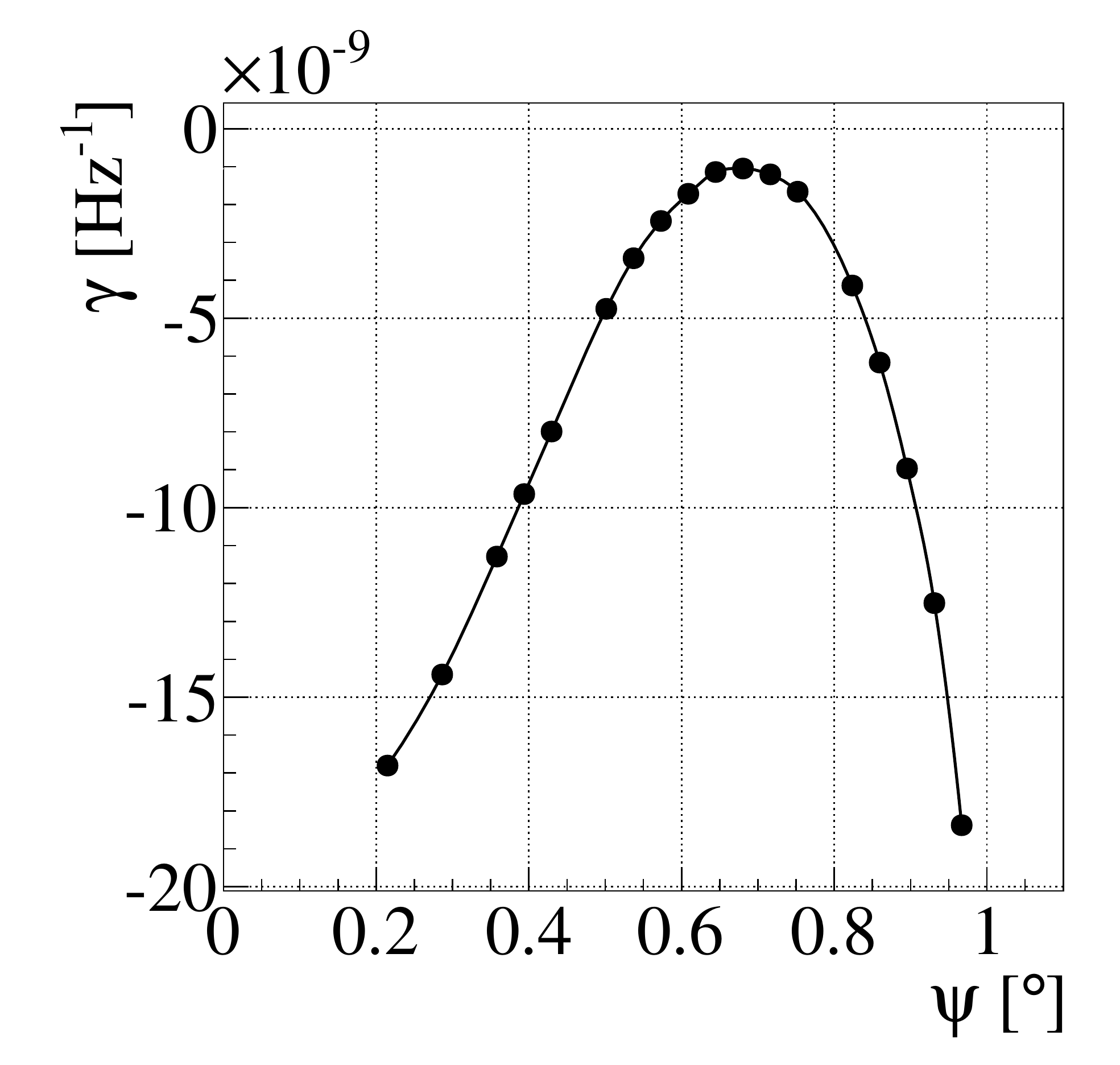}
}
\caption{Left: The amplitude at 300\,MHz, $A$, as a function of the off-axis angle $\psi$ of the location of the  payload. Right: The spectral slope $\gamma$ as a function of off-axis angle $\psi$ of the location of the payload. These results are obtained from a simulated cosmic ray with a 70.5$^{\circ}$ zenith angle and an energy of $10^{18.6}$\,eV. Note, the error bars are smaller than the marker size. The Cherenkov-angle $\psi_c$ is at the off-axis angle where $\gamma$ and $A$ reach their maximum value.}
\label{fig:PsiExamples}
\end{center}
\end{figure}
In Figure \ref{fig:PsiExamples}, we show an example of the amplitude distribution and spectral slope as a function of off-axis angle for payload locations along a line perpendicular to the shower axis. Both distributions peak at the location of the Cherenkov angle $\psi_c$.  As we move away from the Cherenkov cone, we loose power at the higher frequencies and the spectra become steeper. At some point we hit the noisy, mostly incoherent part of
the spectrum, which jeopardizes the fitting of meaningful spectrum slopes.\par
\begin{figure}
\begin{center}
{
\includegraphics[width=0.45\textwidth]{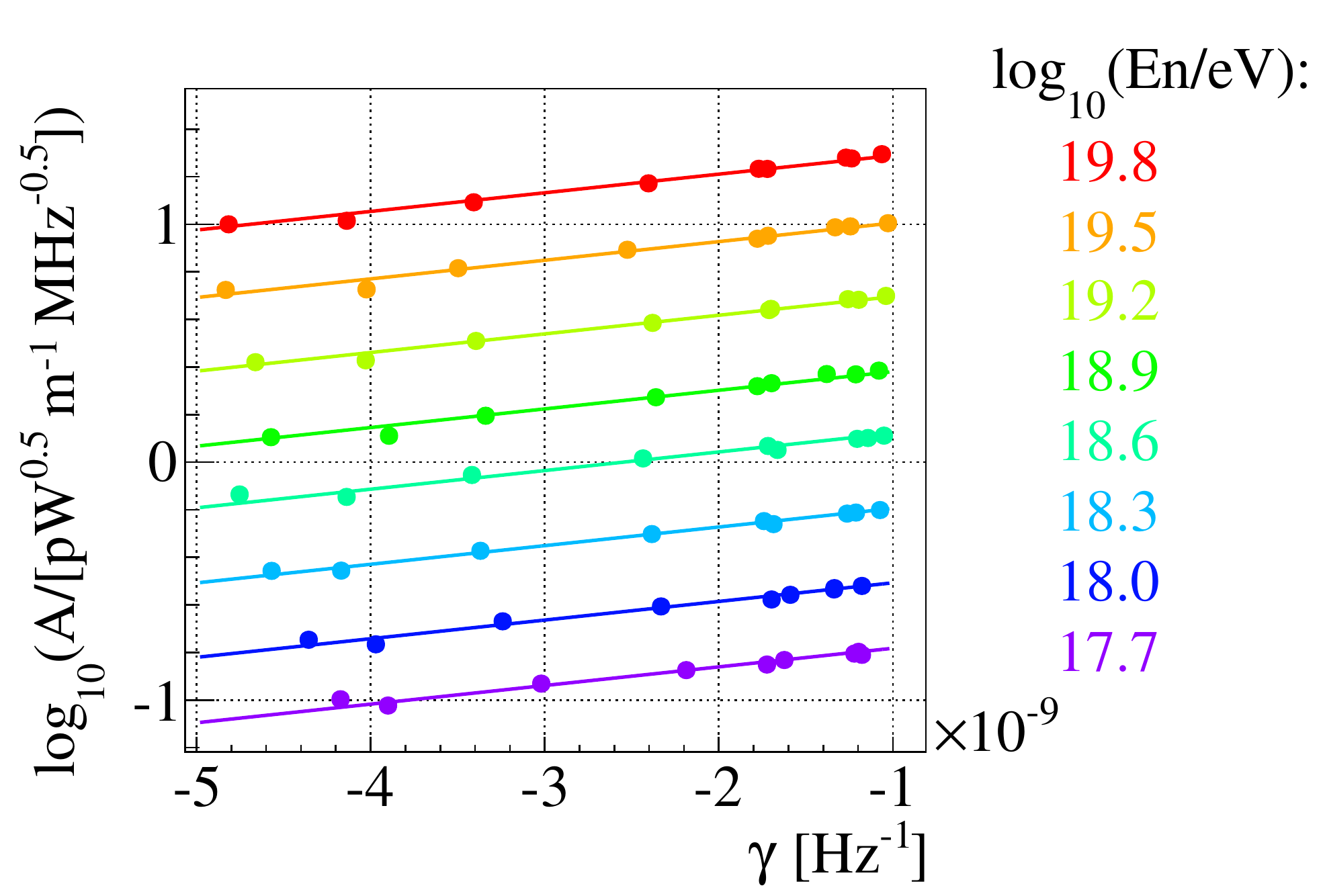} 
\includegraphics[width=0.45\textwidth]{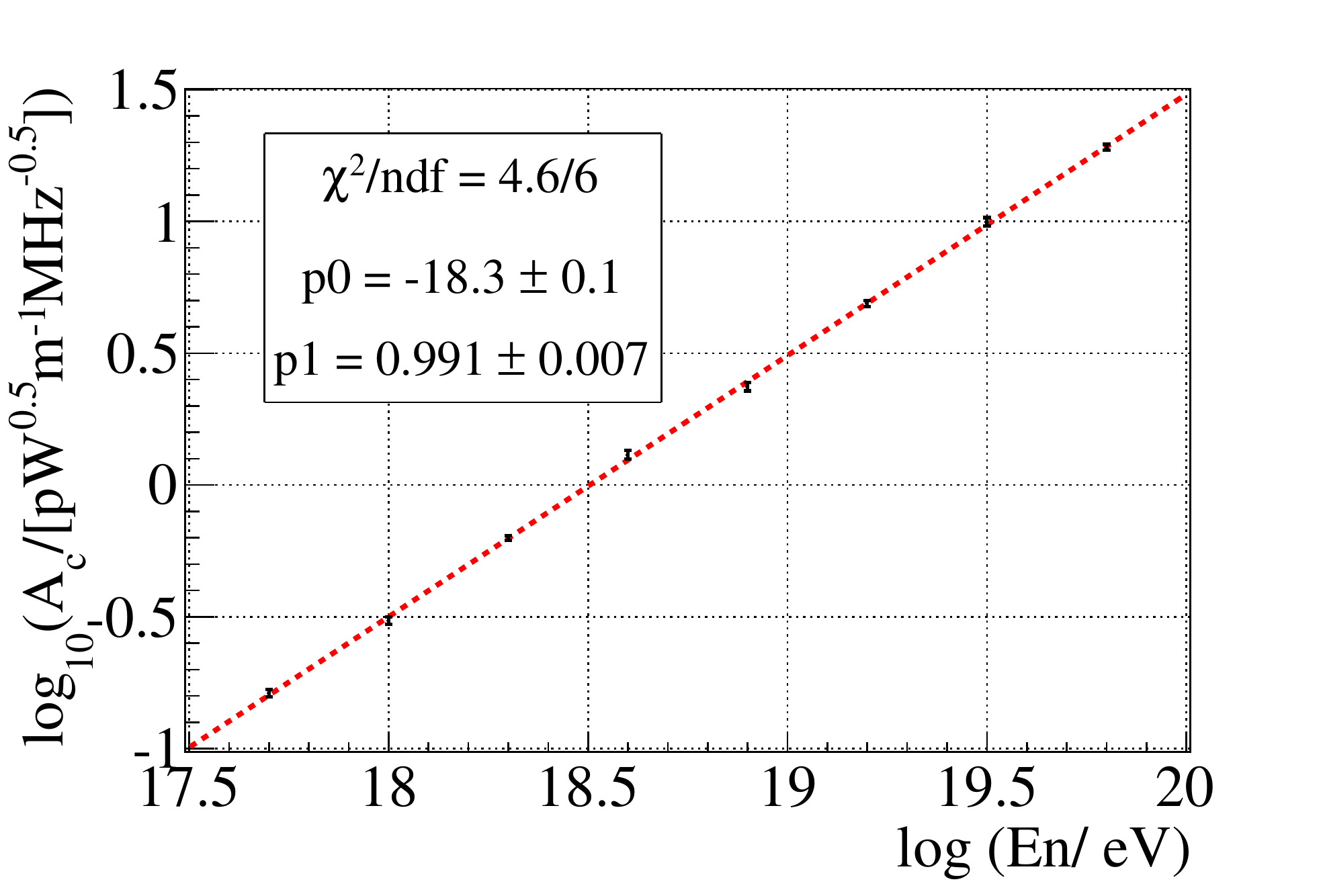}
}
\caption{Left: The amplitude $A$ as a function of the slope $\gamma$ of the amplitude spectra obtained from electric field calculations near the Cherenkov cone for a cosmic ray with zenith angle of 70.5$^{\circ}$. The Cherenkov cone is reached where $\gamma$ and $A$ are at a maximum.  Each marker corresponds to a different off-axis angle. We show this dependency for air showers induced by cosmic rays of different energies. Right: The relation between the amplitude on the Cherenkov cone and the energy of the primary cosmic-ray particle are shown (see the text for detailed explanation). }
\label{fig:RelationSlopeAmplitude}
\end{center}
\end{figure}

 
In the left panel of Figure \ref{fig:RelationSlopeAmplitude}, we show the correlation between $\gamma$ and $A$ from Figure \ref{fig:PsiExamples} in the region near the Cherenkov cone. The color code shows the result of repeating the air shower simulations with different cosmic-ray energy. At different energies the markers can be on slightly different locations because of differences in shower maximum (see Section \ref{sec:corrections} for more details)
We observe that the relation between spectral slope and its amplitude is remarkably simple and can be described by a linear relation:
\begin{equation}
\log_{10}(A) = \log_{10}(A_c) + b(\gamma - \gamma_c),
\label{eq:AAc}
\end{equation}
in which $A_c$ gives the amplitude at 300\,MHz on the Cherenkov cone, $\gamma_c$ gives the spectral slope on the Cherenkov cone, and parameter $b$ describes the dependence of $A$ on $\gamma$. This means that if we determine $b$ and $\gamma_c$ from the simulations, we can derive for each measured spectral slope, $\gamma$, and amplitude, $A$, the corresponding amplitude on the Cherenkov cone, $A_c$. The value of $\gamma_c$ is determined as the flattest frequency spectrum from the set of simulations at different off-axis angles. There is no  significant dependence of $\gamma_c$ on energy, therefore we average it over the different energies. Parameter $b$ is used as a free fit parameter in Equation (\ref{eq:AAc}).  It is clear from Figure \ref{fig:RelationSlopeAmplitude} that it is quite independent of energy. We thus average its value over the different energies considered.

We use the values of $b$ and $\gamma_c$ to calculate for each simulation point in the left panel of Figure \ref{fig:RelationSlopeAmplitude} the value of $A_c$. The right panel of Figure \ref{fig:RelationSlopeAmplitude} shows the dependence of the obtained mean value of $A_c$  as a function of cosmic-ray energy (markers). The (tiny) error bars represent the root mean square of the $A_c$ distribution. These values are fitted with a linear relationship 
\begin{equation}
\log_{10}\left(\frac{A_c}{\rm [pW^{0.5}~m^{-1}~MHz^{-0.5}]}\right) = 
p_0 + p_1 \times \log_{10}\left(\frac{En}{\text{eV}}\right)
\end{equation}
with $p_0=-18.3\pm0.1$ and $p_1=0.991\pm0.007$ (statistical uncertainties from the fit). This linear relationship is used as the cosmic-ray-energy calibration curve for this particular event. The fit results show that the obtained values of $A_c$ are directly proportional to the energy of the cosmic-ray particle.\par 
For each cosmic-ray event, we use the measured values of $\gamma$ and $A$ to calculate the value of $A_c$ using the calibration constants $b$ and $\gamma_c$ obtained from simulations. Using the value of $A_c$ we estimate the cosmic-ray energy from the calibration curve generated specifically for the event.\par
It must be noted that the simple relation between $\gamma$ and $A$ in Equation (\ref{eq:AAc}) starts to break down as one moves further away from the Cherenkov cone. However, the steepest spectral slope obtained from the cosmic-ray observations is -4.0$\times10^{-9}$\,Hz$^{-1}$, which is well in the range where Equation (\ref{eq:AAc}) is a good approximation. As an example, there is still a good agreement of the markers with the linear fit at $\gamma\sim 4\times10^{-9}$ as shown in Figure \ref{fig:RelationSlopeAmplitude} (left) .

\subsection{Reflection coefficients and sources of uncertainty.}
\label{sec:corrections}
Although the ZHAireS simulations are tailored to calculate reflected radio signals for a high altitude balloon, there are still two effects that will alter the electric field observed after the reflection and these are taken into account in this analysis: 
\begin{description} 
\item[Defocusing due to curvature of the Earth:] Since the Earth's surface is curved, parallel incident rays will diverge after reflection. This means that the amount of power-per-area will decrease after the reflection. This decrease depends only on the incident angle of the radiation. It becomes more significant near the horizon where the incident angles of the radiation are large with respect to the surface. To calculate the loss we used the analytical model from \cite{andres} and validated its accuracy with the ray tracing algorithm explained in \cite{ref:ZHSreflex}.
  \item[Loss of coherence due to surface roughness: ] If the roughness of a reflecting surface is of a similar or larger size than the wavelength of the radiation, a reflection will disturb the wavefront and there will be a loss of coherence. We estimated the loss of coherence by performing a physical optics calculation over a surface using the method described in \cite{ref:SWORD}. The model of surface roughness used in the calculation is based on a parameterization \cite{ref:radarscatter} of surface roughness measurements performed over length scales between 0 and 120\,m at the Antarctic Taylor Dome station. The parameterization is rescaled to reflect the roughness at other locations using the digital elevation model from \cite{RADARSAT} which provides elevation information on a 200\,m spaced grid. From evaluating our physical optics calculation it turns out that we are mostly interested in surface roughness on length scales between 10 and 30\,m. Therefore, we assign a significant uncertainty to the roughness correction. \par
 A rougher surface with respect to the currently used values induces a smaller amplitude and a steeper spectrum. The smaller amplitude implies that the energy of the event would be underestimated, but on the other hand the steeper spectrum implies that the off-axis angle would be reconstructed closer to the Cherenkov angle resulting in an overestimate of the energy. These two effects almost cancel each other out and the resulting uncertainty on the energy due to the uncertainty in roughness is the range 1\%-6\%.\par
However, surface roughness becomes more important when simulating the ANITA-I flight as both the steepening of the amplitude spectrum and the reduced amplitude will reduce the likelihood of a trigger at the ANITA-I instrument.
\end{description}
\begin{figure}
\begin{center}
{
\includegraphics[width=0.7\textwidth]{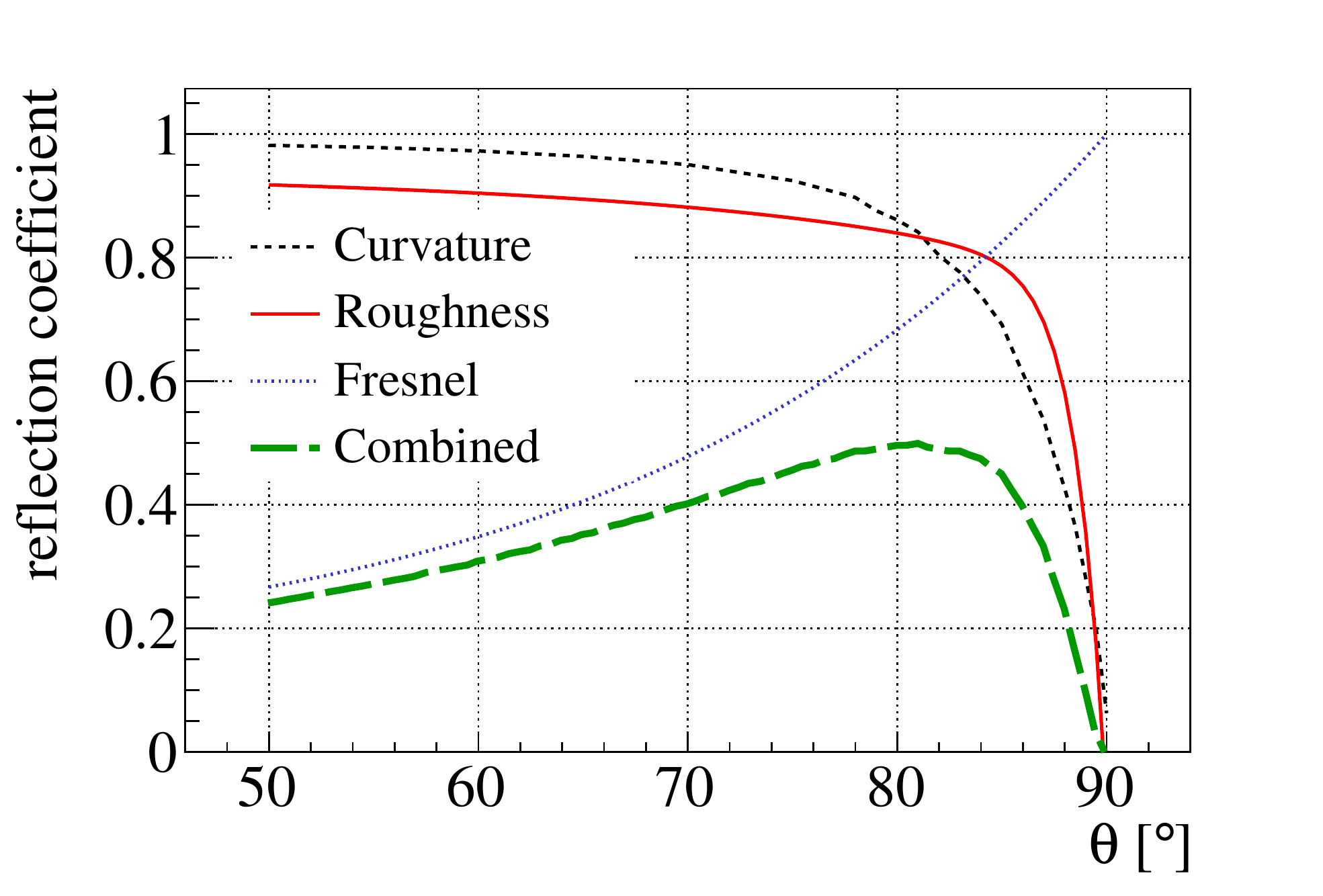} 

}
\caption{The various contributions to the reflection coefficient as a function of incident angle. The green dashed line shows the scaling factor for the combination of all the contributions. For illustration purposes, the surface roughness factor is taken at 650\,MHz. In the full calculations the frequency dependency of this factor is taken into account as described in Appendix A.}
\label{fig:scalingFactors}
\end{center}
\end{figure}

These two effects, in combination with the Fresnel coefficients, give the total scaling factor for simulated electric fields. The Fresnel coefficients are already included in the simulation package. In Figure \ref{fig:scalingFactors} we show how the scaling factors depend on the incident angle of the radiation.\par
There are several parameters that influence the cosmic-ray energy estimation  that cannot be constrained from the observations made by ANITA. As a result, these parameters contribute to the uncertainty in the cosmic-ray energy estimate. Below we describe each of these contributions and how we treated them to assess their contribution to the uncertainty. 

\begin{description}
  \item[Ambiguity in the direction of the air shower:] The radio signal is emitted in a cone-like beam around the shower axis. Therefore, there is an offset between the direction of the radiation and the direction of the shower axis. 
 There are also two effects that cause asymmetries in the radio beam pattern that depend on which side of the shower axis the detector is located: Firstly, there is a radially-polarized contribution,with respect to the air shower axis, due to charge-excess radiation (Askaryan radiation) that interferes with the dominant geomagnetic radiation \cite{ref:AugePol}. At the zenith angle range of the air showers and the frequency range considered here, this interference alters the observed emission only at the percent level. Secondly, a stronger source of asymmetry is due to whether the zenith angle of the shower axis was smaller or larger than the observed incident angle of the radiation ($\theta$). The altitude at which the air shower develops becomes more strongly dependent on the zenith angle of the shower axis when the zenith angle increases. The zenith angle will influence the distance from the detector to the emitting region and the amount of radiation emitted by the air shower. \par
 To sample these asymmetries associated with the payload location with respect to the air shower axis we pick four air shower offset directions. If we describe the direction of the radiation by the incident angle on the surface, $\theta$, and azimuthal angle, $\phi$, then the following offsets were chosen for the direction of the shower axis: ($\theta + \psi_c,\phi$),($\theta - \psi_c,\phi$),($\theta,\phi- \psi_c$) and ($\theta ,\phi+ \psi_c$). The payload locations are distributed along lines that extend from the location of the offset through the location of observed radiation to an off-axis angle of 2.4 times $\psi_c$. We estimate the calibration constants $\gamma_c$ and $b$ and calibration lines (as in Figure \ref{fig:RelationSlopeAmplitude}) on each of these four air showers individually to reconstruct the energy. We report the average value as the energy of the cosmic ray and use the spread in these values as an uncertainty on the cosmic-ray energy. The uncertainty on the individual cosmic-ray energy varies from 4\% to 15\% and are listed in Table \ref{tab:EnergyInduvidual}.\par
 The uncertainty on angular reconstruction is small $\delta \theta \approx 0.2^{\circ}$ and $\delta \phi \approx 0.8^{\circ}$ and compared to the offsets discussed above the resulting uncertainty on the energy is negligible.\par 
  \item[Uncertainty on the atmospheric depth of air shower development: ] Most of the radiation is expected to come from near the region where the air shower reaches its maximum number of particles, usually referred to as an atmospheric depth $X_{\text{max}}$ expressed in g\,cm$^{-2}$. The location of $X_{\text{max}}$ fluctuates from shower to shower and the distribution of $X_{\text{max}}$ depends on the energy of the cosmic ray. The most accurate measurement of the $X_{\text{max}}$ distributions in the energy range of interest are performed by the Pierre Auger Observatory \cite{ref:AugerXmax}. We force the air shower simulations used for the energy estimation of individual events to be near ($\sim$5 g/cm$^{2}$) the observed mean $\langle X_{\text{max}} \rangle$ as measured by the Pierre Auger Observatory. In this way the simulations are tuned toward observations and not the choice of interaction model.\par
To estimate the uncertainty on the energy reconstruction due to the spread of $X_{\text{max}}$ we generated a set of 50 proton air shower simulations to obtain a distribution in $X_{\text{max}}$ values (without forcing the simulations to be near $\langle X_{\text{max}} \rangle$  observed at Pierre Auger Observatory). By estimating the  the cosmic-ray energy for each of these showers individually we obtain a distribution of reconstructed energies. The spread of this distribution is used  as the uncertainty on the energy estimation due to variation in shower development and is of the order of 4\%. The variation in $X_{\text{max}}$ is largest for proton induced air showers, therefore the uncertainty of 4\% can be considered as an upper limit. Note, the uncertainty on the energy of only 4\% due to proton $X_{\text{max}}$ fluctuations indicates that there is little dependency (and sensitivity) on the exact depth of shower maximum whatsoever. 
 \item[Variation in atmospheric refractive index:] From atmospheric measurements made throughout the austral summer, we model the atmospheric refractive index. To the average atmospheric refractive index profile we fit an exponential function that is used in the simulation of the radio signals. The deviations from this fit are $\sim3$\% at an altitude relevant for air showers. To estimate the impact of these deviations on the energy estimation, we varied the refractive index model in the simulations and evaluated the impact on $A_c$. This resulted in uncertainty on the cosmic-ray energy estimates that range from 4\% to 9\%.
  \item[Variation of the refractive index of the snow surface:] The Fresnel coefficients depend directly on the dielectric properties of the surface. For the Fresnel coefficients used in our analysis an average density of the firn surface is assumed which leads to a refractive index of 1.35. A layer of fresh snow reduces the average density which can lead to a 10\% lower refractive index. The effect of a 10\% decrease of the refractive index lowers the reflection coefficient by 25\,\% at an incident angle of 57$^{\circ}$ and 8\% at an incident angle of 85$^{\circ}$. Because of the lack of knowledge of the exact refractive index of the snow surface at the location of each reflection, we assign an uncertainty to the Fresnel reflection coefficient corresponding to 10\% deviation in the refractive index.
 \item[Calibration of the ANITA-I instrument:]   The observations have been corrected for instrument response. The instrument response is characterized by measurements of the effective height of the antenna and the impulse response of the system after the antenna. An uncertainty of 1\,dB is assigned to each individual event to take into account the imperfections in the gain corrections per signal chain. In addition, we adopt a 1\,dB (12\%) systematic uncertainty on the energy scale to account for the uncertainty in the method to measure the antenna effective height.
 \item[Uncertainty due the radio simulation package used:] We compared the electric field obtained with the ZHAireS and CoREAS packages for antennas on the ground, for details see Appendix B. Deviations in $A_c$ were found from 5\% up to 30\% depending on the zenith angle of the air shower. We adopt these deviations as an uncertainty in the energy reconstruction per individual event. 
 \item[Uncertainty on parameters derived from the measurements:] The statistical uncertainty on the fit parameters $A$ and $\gamma$ are propagated to the uncertainty on the energy of the cosmic rays. This uncertainty contributes between 6 and 38\% of the cosmic-rays energy.
 \item[Uncertainty on calibration constants derived from the simulations:]  While generating the simulations to obtain the calibration constants we have to a make few pragmatic choices. The simulations have a finite number of locations where the electric field is calculated and the values of $A$ and $\gamma$ are linearly interpolated between these locations. In addition, there is a slight variation in the depth of shower maximum for each different cosmic-ray energy. To estimate the impact on calibration constants $b$ and $\gamma_c$, the spread on these constants are calculated for the different cosmic-ray energies. The spread on these constants are propagated to the uncertainty on the cosmic-ray energy and varies between 2\% and 10\% of the cosmic-ray energy.

\end{description}  

\subsection{Cosmic-ray energy estimation of individual events}
\begin{figure}
\begin{center}
{
\includegraphics[width=0.7\textwidth]{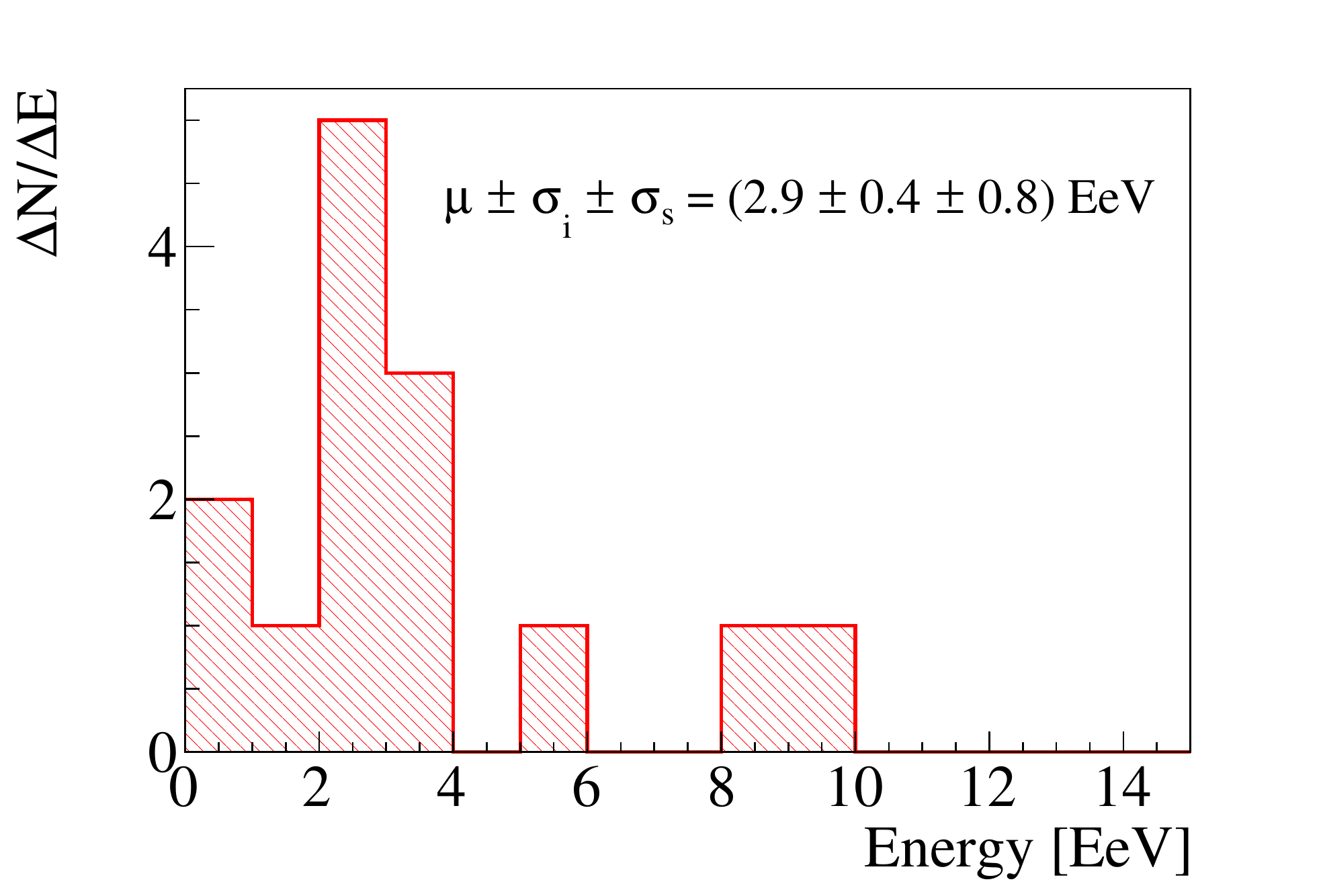} 
}
\caption{Distribution of number of events $\Delta N$ per energy bin $\Delta E$ (with $\Delta E=1$\,EeV) for the 14 cosmic-ray events. Shown are the weighted mean $\mu$, the uncertainties on it due to uncertainties on the individual events $\sigma_i$  and the uncertainty on it due to the absolute scale $\sigma_s$. The quoted uncertainties represent 95\% confidence intervals.}
\label{fig:EnergyDistribution}
\end{center}
\end{figure}

We applied the method and corrections discussed in the previous sections to the recorded signals to obtain the energy of the individual cosmic rays. The measured  cosmic-ray energy distribution is shown in Figure \ref{fig:EnergyDistribution} and the individual energy estimates are listed in Table \ref{tab:EnergyInduvidual}.\par 

\begin{table}[htdp]
\begin{center}
\begin{tabular}{|c|c|c|c|c|c| }
\hline
Event & Energy &	Uncertainty 		&Unc. Amb. dir.    &$\psi_c$&$\Delta\psi_c $ \\ 
\#        &  [EeV] 	&  [\%]				&  [\%]			  &$[^{\circ}]$& $[^{\circ}]$  \\\hline
1*         & 	2.1	& 43 				&  	14		&0.43		&0  \\
2*         & 	3.1	& 49				&    	12		& 0.52		&0  \\
3         & 	2.3	& 30				&	4		&  0.78		&0.13 \\
4         & 	3.1	& 28				&    	5		& 0.78		&0.15 \\
5         & 	9.0	& 27				&  	5		&  0.80		&0.16 \\
6         & 	2.1	& 25				 &   	6		& 0.73		&0.06 \\
7         & 	1.0	& 27				&	4		&  0.62		&0.07 \\
8         & 	9.9 	& 55				&	15		&  0.93		&0.20 \\
9         & 	1.0	  & 29				&	5		&  0.63		&0.05 \\
10       & 	2.6	& 24				& 	4		&  0.55		&0.09 \\
11       &  	3.9	& 24				&	5		&  0.70		&0.08 \\
12       & 	1.9	& 28				&	6		&  0.55		&0.03 \\
13       & 	5.6	& 34				&	12		&  0.49		&0.12 \\
14       & 	2.7	& 23				&	3		&  0.56		&0.07  \\\hline
 \end{tabular} 
\caption{Energy estimate on the individual events and the total uncertainty on it due to the several sources of uncertainty listed in Section \ref{sec:corrections}.  The column labelled Unc. Amb. dir.  gives the fraction of the uncertainty due to the effect discussed in {\bf{Ambiguity in the direction of the air shower}} in Section 4.4. Also listed are the off-axis angle of the Cherenkov cone and the reconstructed offset with respect to the Cherenkov angle. The two events marked with an `*'  had flatter observed spectral slopes than obtained from the simulations and are therefore assumed to be on the Cherenkov cone (see the text for more details).}
\label{tab:EnergyInduvidual}
\end{center}
\end{table}
From Table \ref{tab:EnergyInduvidual} we calculate the weighted mean energy of the cosmic-ray sample to be 2.9\,EeV. The weights are set by  quadratic sum of contributions to the uncertainty (as listed in Table \ref{tab:EnergyInduvidual})  that are expected to cause random event-to-event fluctuations in the energy estimate. On the mean energy we estimate an 95\% confidence interval $\sigma_i$ due to these uncertainties of 0.4\,EeV. In addition to this uncertainty, we estimate an  95\% confidence interval $\sigma_s$ due to sources of uncertainty that could systematically influence the overall energy scale to be $0.8$\,EeV.\par
The two events in Table \ref{tab:EnergyInduvidual} that are marked with an `*'  had flatter observed spectral slopes than obtained from the simulations. In the case of event 1, the flattest spectral slope from simulations is within the uncertainty of the spectral slope compatible with the observation. Therefore, it is reconstructed on the Cherenkov cone. This was not the case for event 2. The roughness rescaling to obtain a roughness model at the location of the reflection is the largest among all the events. However, this large rescaling factor also has a large uncertainty on it. Simulating different surface models with rescaling factors distributed to reflect the uncertainty on the used rescaling factor results in simulated spectral slopes that are compatible with the observed spectral slope in 30\% of the cases. Therefore we decided to reconstruct the energy as though it was reconstructed on the Cherenkov cone. The deviation in $A_c$ between the largest roughness scaling parameter that resulted in a reconstructable energy and the nominal used roughness scaling factor is used to set an additional uncertainty on the energy reconstruction of 30\%. This event underlines the importance of having accurate values for the local surface roughness.\par 
In addition to the cosmic-ray energy estimations, we also present the off-axis angle of the Cherenkov cone $\psi_c$ in Table \ref{tab:EnergyInduvidual} as obtained from the simulations. From the spectral slope distributions obtained from simulations, e.g., the right panel of Figure \ref{fig:PsiExamples}, we obtain the two off-axis angles where the measurement intercepts the distribution. In the last column of Table \ref{tab:EnergyInduvidual}, we report the average angular offset of these two angles with respect to the Cherenkov cone, $\Delta \psi_c$.\par 

\section{Simulating the ANITA flight}
\subsection{Full Monte Carlo simulation of the ANITA flight}
\begin{figure}
\begin{center}
{
\includegraphics[width=0.7\textwidth]{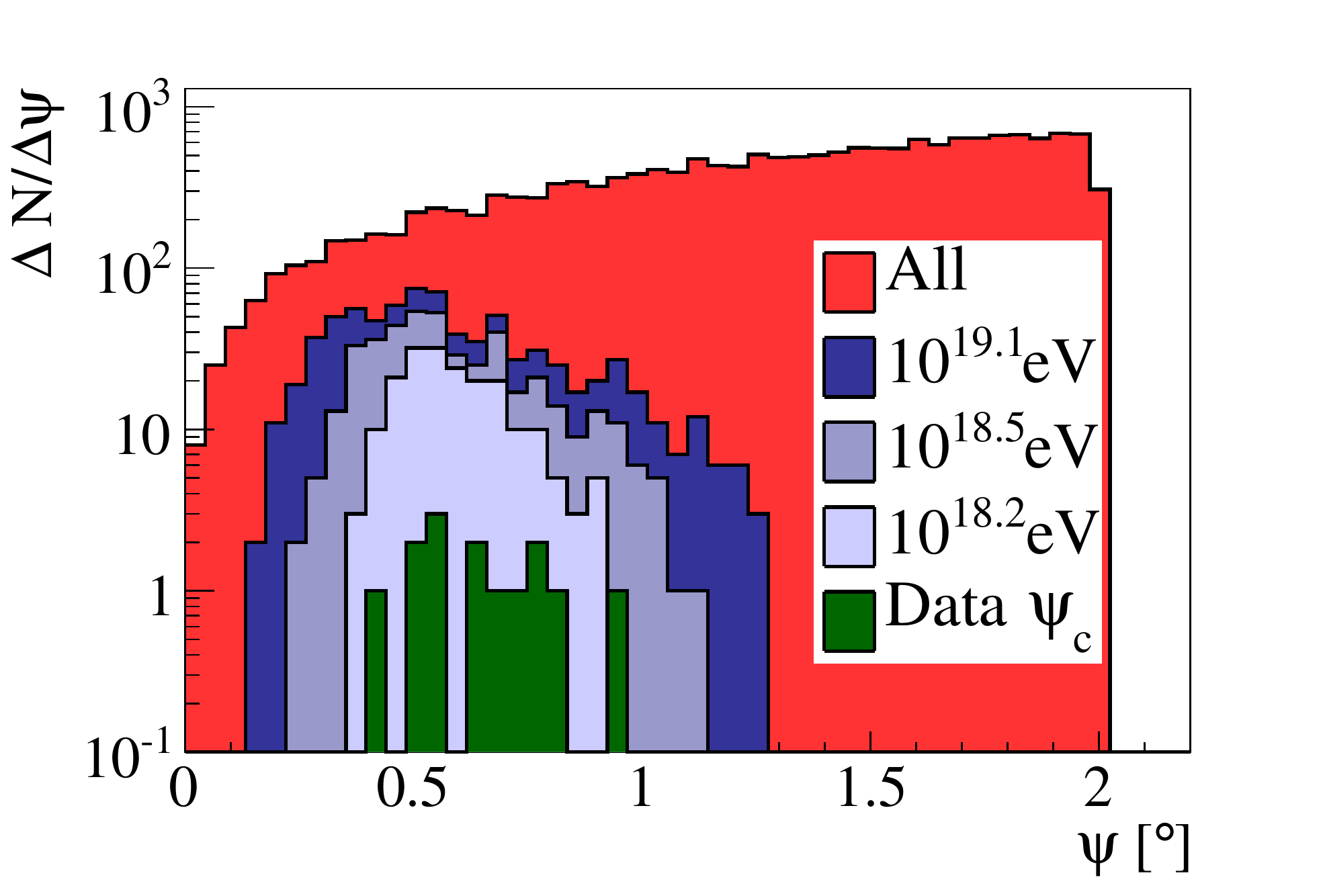} 
}
\caption{The distribution of the number of events $\Delta N$ per off-axis bin $\Delta\psi$ (with $\Delta\psi = 0.044^{\circ}$) obtained from the flight simulation. We compare the off-axis distribution of all simulated events to the distributions of events that resulted in a trigger. The distributions are shown for cosmic rays simulated with different energies. In addition we show the distribution of values of $\psi_c$  for the measurements (Table 2)}
\label{fig:psidist}
\end{center}
\end{figure}

In order to estimate the exposure, we need to calculate the trigger efficiency of ANITA to an isotropic cosmic-ray flux. The method to calculate the exposure is partially inspired by the calculations performed in \cite{ref:priv} and the detector simulation is based on the flight performance as reported in \cite{ref:AnitaExp}. Below we describe the strategy of the flight simulation in a (semi-) sequential order:
\begin{enumerate}
  \item The payload is located at a fixed location with an altitude of 36\,km above sea level, which corresponds to the average altitude during the flight.
  \item The Antarctic surface is chosen at altitude of 2\,km above sea level, which corresponds to the average height of the Antarctic ice sheet overlooked by ANITA during the flight.
  \item Seeds for cosmic rays are generated in the following way: We choose a surface 200\,km above sea level parallel to the surface of the Earth. The size of the surface (4.5$\times 10^{7}$\,km$^2$) extends slightly over the horizon that ANITA can see from 36\,km altitude. On this surface seed locations for cosmic rays are generated by drawing from a uniform distribution function. To each seed a direction is assigned by drawing a direction from an isotropic distribution. In total 6.7$\times 10^8$ seeds are generated.
  \item A cosmic ray is tracked into the atmosphere and the location is calculated where it reaches a slant depth of 735\,g\,cm$^{-2}$, which corresponds roughly to the average value of $X_{\text{max}}$ above 10$^{18}$\,eV. This location is used to make a decision: if the off-axis angle with the location of the payload is less than 2$^{\circ}$ we will start the full simulation of the particle shower and the radio emission. If it is larger than 2$^{\circ}$, it is unlikely that it will trigger the instrument and therefore it will be rejected (see Figure \ref{fig:psidist} for validation of this cut). The number of cosmic rays passing this cut is 16968. 
  \item The full simulations are run at a fixed cosmic-ray energy of 10$^{18.4}$\,eV and protons as a primary cosmic-ray particle. To explore the energy dependency of the exposure, the radio signal is scaled proportional to the cosmic-ray energy afterward (this linear scaling is a valid assumption see for example Figure \ref{fig:RelationSlopeAmplitude} (right)). The geomagnetic field input parameters of the simulations are chosen such that they mimic the geomagnetic field configurations along the flight path. The orientation of the shower direction with respect to the geomagnetic field is the dominant factor that determines the polarization of the signal. By varying the geomagnetic field configurations and the incoming directions the polarization dependency of the trigger efficiency is taken into account. 
  \item The roughness corrections are applied to the resulting electric field obtained from the simulations. Several scenarios for the reflection properties of the surface are simulated to estimate the influence of the surface properties on the trigger efficiency. 
  \item The electric field at the payload, corrected for reflection effects, is propagated through the antenna model and the analog chain taking into account the polarization of the signal and the antennas. 
  \item To the signal, several noise contributions are added. The noise seen in the field of view of the antenna is the combination of the noise temperature of the ice and the sky  which is roughly 190\,K over the full frequency band. Narrowband emission between 220 and 290\,MHz is also present in the field of view, arising from satellites. This emission was observed toward the north side of the payload and therefore their contributions are simulated into the north facing antennas. The power is set to match observations during the flight. In addition to the noise seen by the antennas, the noise generated by the full trigger chain is simulated based on measurements performed before flight and is about 140\,K over the frequency band. 
  \item The combination of signal and noise is fed into the trigger simulation. The trigger simulation is setup with typical thresholds and multiplicity of frequency bands (minimal three out of eight) and antenna multiplicity (four or more adjacent antennas) as used during flight. The thresholds are set such that they correspond to a fix rate of threshold crossings per single antenna, therefore it depends on the amount of noise seen by an antenna. We ran the simulation once with the nominal noise setup (as described in the previous item), once reflecting a more noisy environment (near a base) and once reflecting a quiet region of the flight. 
 \item The fraction of accepted cosmic rays (not rejected in step 4 or 9)  with respect to the total number of cosmic rays simulated in step 3 is used to calculate the effective aperture of the ANITA experiment. Using the total life time of the experiment of 17.25 days the exposure is estimated.
\end{enumerate}

\subsection{Exposure and cosmic-ray flux calculation}
\begin{figure}
\begin{center}
{
\includegraphics[width=0.7\textwidth]{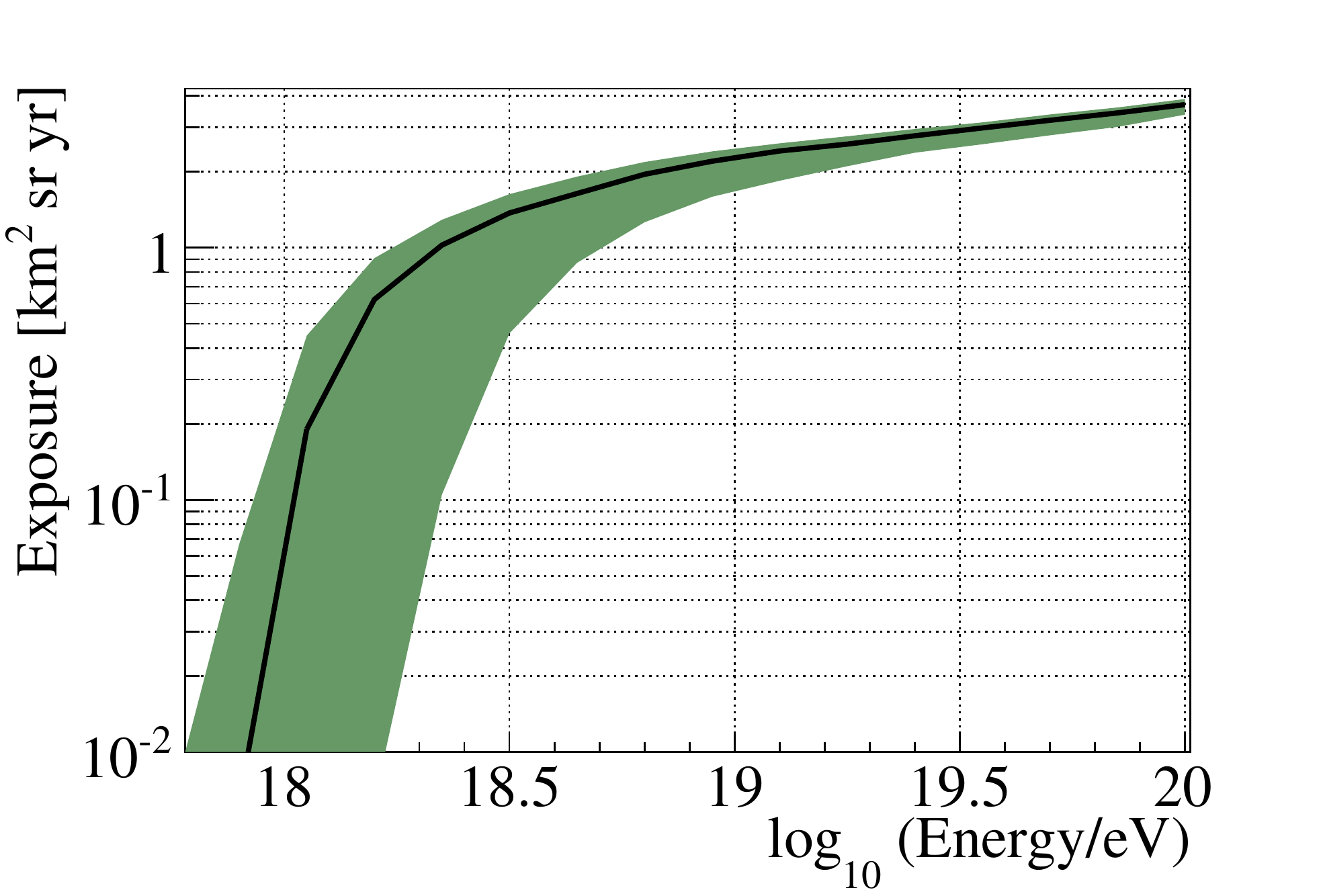} 
\includegraphics[width=0.7\textwidth]{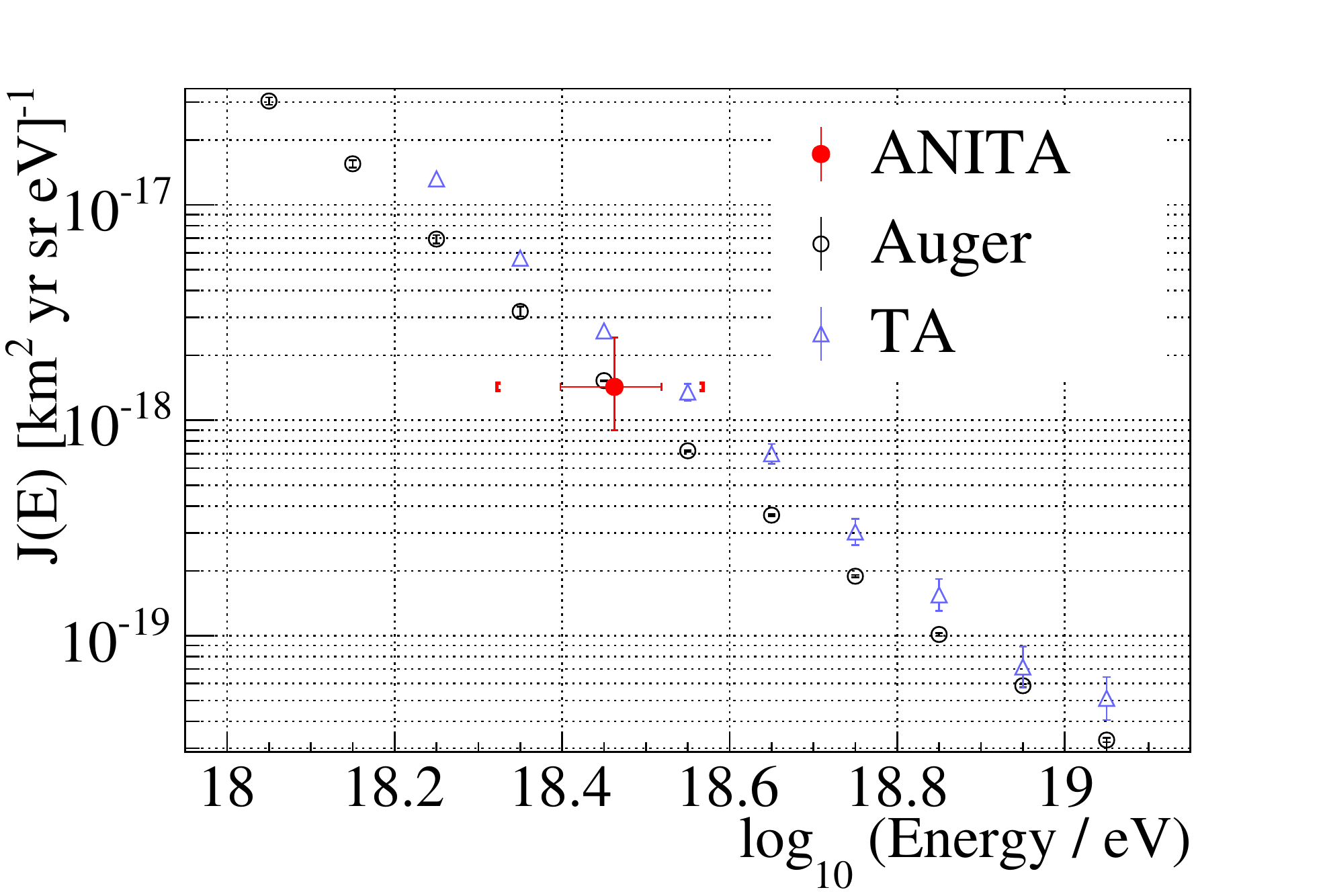} 
}
\caption{Top: ANITA-I's exposure to UHECRs, derived from Monte Carlo simulations. In additional to the nominal values (black line)  for the reflective properties of the ice we ran several scenarios varying the properties of the reflective surface. All the scenarios are contained within the green shaded area (see the text for more details).  Bottom: Comparison between the UHECR flux as observed by ANITA-I (see the text for details),  the Pierre Auger Observatory \cite{ref:AugerICRC} and the Telescope Array \cite{ref:TA}.}
\label{fig:exposure}
\end{center}
\end{figure}

The trigger efficiency from the flight simulation determines the total exposure of the flight, which is displayed in the top panel of Figure \ref{fig:exposure}.
The energy scale at which the exposure grows rapidly is approximately the same energy scale as the average energy from the individual events. The fast rise in exposure below 10$^{18.5}$\,eV basically reflects the energy regime where some of the air showers start to be above the detection threshold. At a higher energy, the growth in exposure slows down and the steady increase is due to the increasing possibility of detecting events further away from the Cherenkov cone, as seen in Figure~\ref{fig:psidist}.\par
In the bottom panel of Figure \ref{fig:exposure} we compare the cosmic-ray energy flux $J(E)$ as observed by the Pierre Auger Observatory and the Telescope Array to the flux observed by ANITA-I. Due to the small number of events measured by ANITA, we decided to calculate the flux using a single bin that contains the whole energy range of the observations and use the average reconstructed energy to set the energy scale. The vertical error bar is the quadratically summed combination of the Poisson uncertainty on the number of events and uncertainty on the exposure.\par
To assign an uncertainty to the exposure we performed the full flight simulation in several scenarios by varying the roughness model (twice as rough or half as rough as at Taylor Dome), the background model (quiet conditions and noisy conditions) and the average refractive index of the ice (n = 1.31 and n = 1.35).  Increasing the surface roughness by a factor of two results in a significant decrease in the exposure by a factor 0.3 at $2.9\times10^{18}$\,eV, while reducing it by a factor of 0.5  increases the exposure only by a factor 1.05.  Changing the refractive index of the ice from 1.35 to 1.31 leads to a 3\% increase of the exposure at $2.9\times10^{18}$\,eV. Using a background model that reflects quiet conditions increases the exposure by 6\% while more noisy conditions reduce the exposure by 11\% at $2.9\times10^{18}$\,eV. To be conservative, we used the two most extreme combinations of these scenarios  as an uncertainty on the exposure.\par 
Using this approach we find that the observed cosmic-ray flux by ANITA at 2.9 EeV is 1.4$^{+1.0}_{-0.5}$\,km$^{-2}$\,yr$^{-1}$sr$^{-1}$eV$^{-1}$. The horizontal error bars are given by the 95\% confidence interval $\sigma_i$ due to the uncertainties on the individual events of 0.4\,EeV while the brackets indicate the  95\% confidence interval  $\sigma_s$ due to the uncertainty on the overall scale of the radio signal of $0.8$\,EeV.\par

\subsection{Cosmic-ray observations compared to expectations from Monte Carlo simulations}
Using the exposure, we estimate the number of expected cosmic rays via Monte Carlo simulations. To do so, we integrate the parameterized cosmic-ray-energy spectrum, as measured at the Pierre Auger Observatory \cite{ref:AugerICRC} (or at the Telescope Array \cite{ref:TA}), multiplied with the exposure function over the full energy range. The resulting number of expected events is 16 (Pierre Auger spectrum) or 20 (Telescope Array spectrum), compared to the 14 observed events.\par
\begin{figure}
\begin{center}
{
\includegraphics[width=0.6\textwidth]{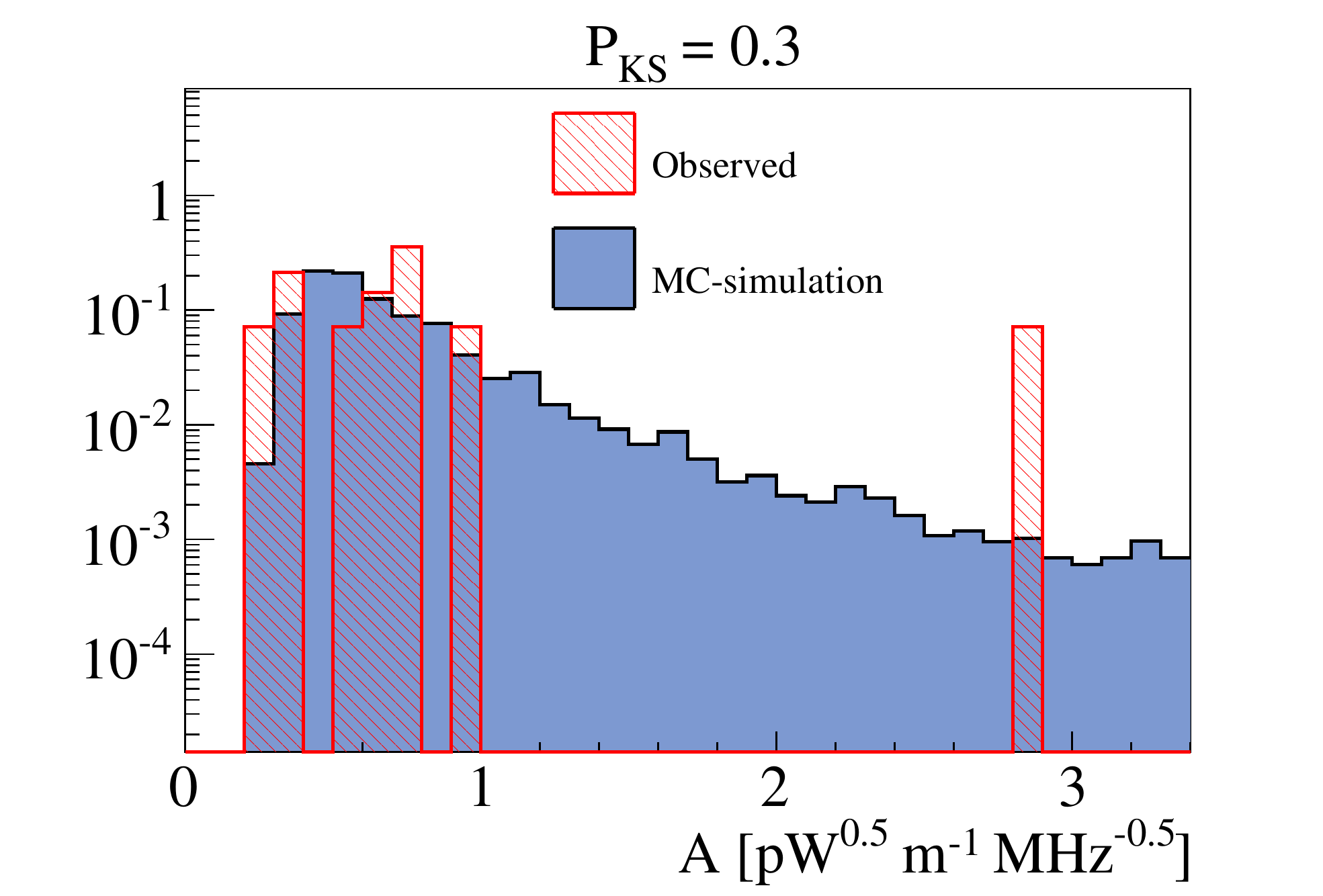} 
\includegraphics[width=0.6\textwidth]{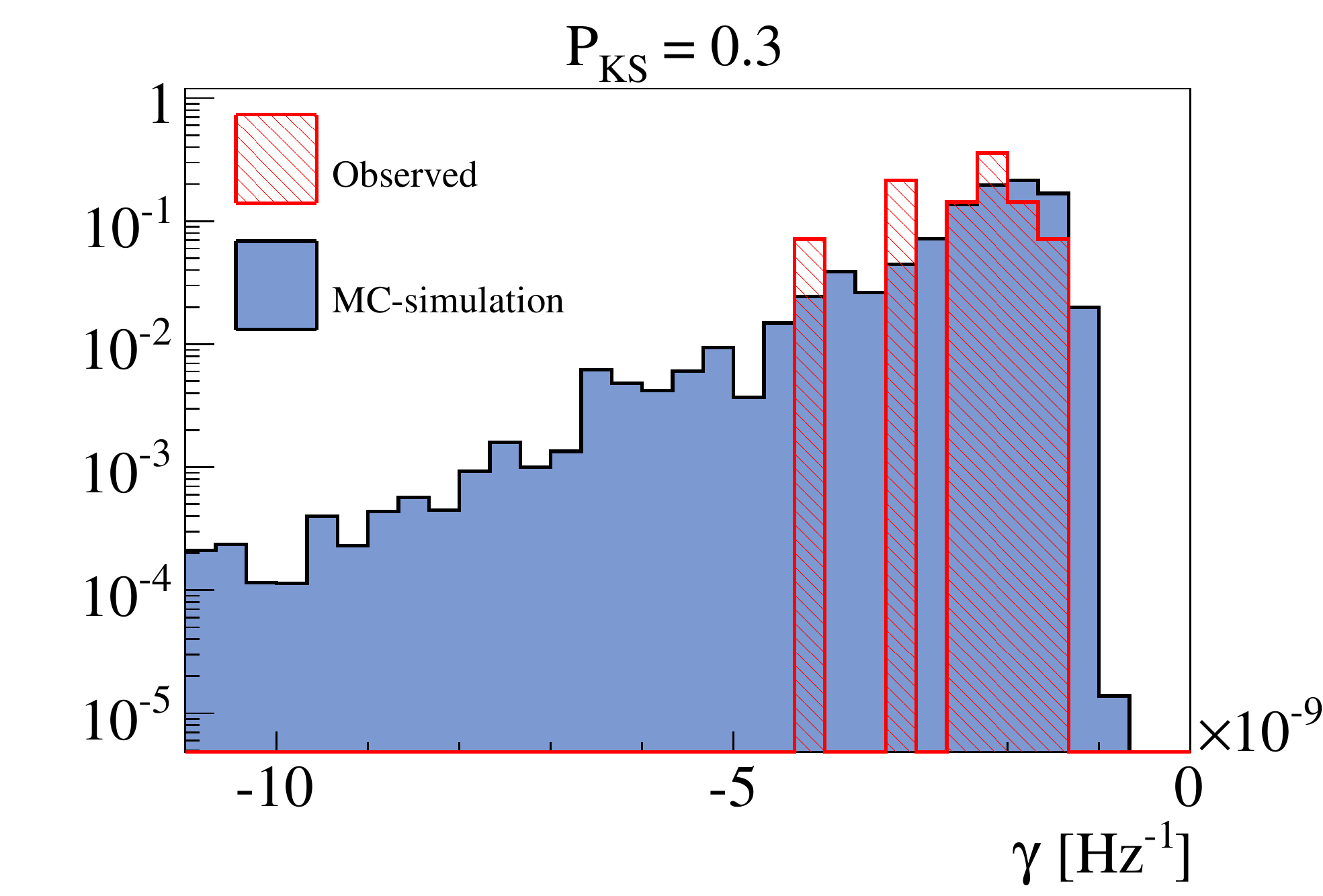} 
\includegraphics[width=0.6\textwidth]{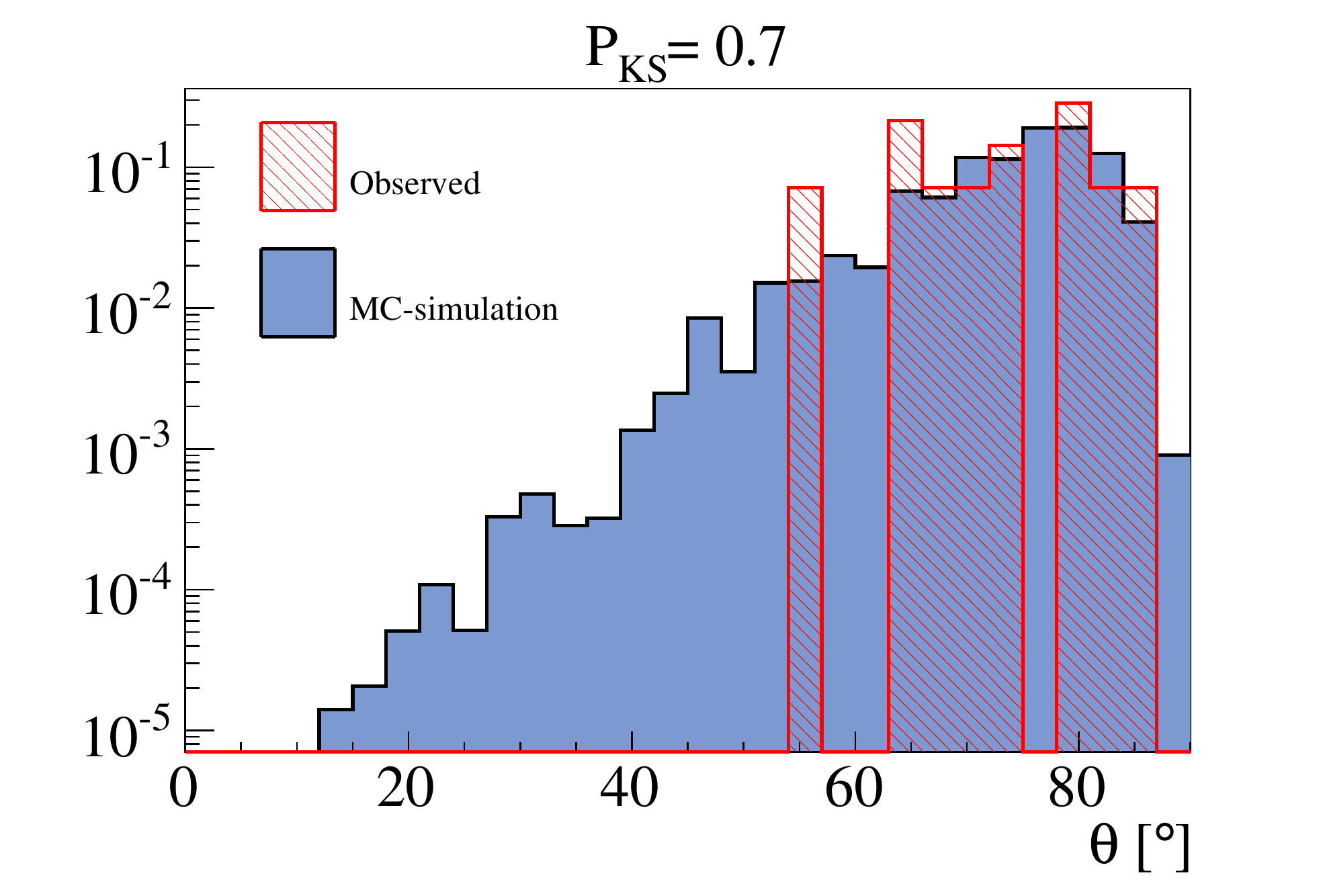} 
}
\caption{Comparison between distribution of parameters obtained from the flight simulation to observations. The bin sizes are chosen of equal size for simulations and observations and the vertical scale is normalized so that integral over each distribution equals 1. Top: Distribution in amplitude $A$. Middle: Distribution in spectral slope $\gamma$. Bottom: Distribution in incident angle on the ice. To the entries in the Monte-Carlo distributions weights are applied to reflect the energy dependence of the cosmic-ray spectrum. We used a weight proportional to the number of events as a function of energy as expected from Pierre Auger Observatories cosmic-ray-energy spectrum.} 
\label{fig:distAmpSlopeZen}
\end{center}
\end{figure}

In addition to the expected total number of events, we can also compare the distributions of several variables obtained from the measurements to those of the flight simulation. We compared the following parameters: $A$, the amplitude of the frequency spectrum of the signal at 300\,MHz, $\gamma$, the spectral slope fitted between 300-1000\,MHz, and $\theta$, the incident angle of the radiation on the ice. The results are given in Figure \ref{fig:distAmpSlopeZen}.
The distributions obtained from the Monte Carlo simulations are weighted as a function of energy by the number of events expected from the cosmic-ray energy spectrum as observed at the Pierre Auger Observatory or the Telescope Array. To test the compatibility of the measured and simulated distributions we normalized them both and perform a Kolmogorov-Smirnov test. In Figure \ref{fig:distAmpSlopeZen}, we report the probability of having the calculated Kolmogorov-Smirnov test statistic or larger, $P_{KS}$, using the Pierre Auger energy spectrum. This indicates that there is statistical agreement between the distributions obtained from simulations and observations. Using the cosmic-ray energy spectrum as observed by the Telescope Array, we find the following probabilities $P_{KS} = 0.4$ for the distributions in $A$,  $P_{KS}=0.5$ for the distributions in $\gamma$ and $P_{KS} = 0.7$ for the distributions in $\theta$.\par

\section{Discussion}
The distribution of observed amplitudes contains one obvious outlier, this event has been identified as number 5 in Table \ref{tab:EnergyInduvidual}. We inspected this event carefully however, and except for its exceptionally large amplitude, the other parameters of this event are in the bulk of their corresponding distributions. Thunderstorm conditions are known to cause enhanced radio signals from air showers \cite{ref:ThunMand,ref:ThunLOPES,ref:ThunLOFAR} and therefore provide a feasible candidate to explain outliers in the amplitude distribution. However, thunderstorm conditions are extremely rare in Antarctica \cite{ref:NASAlighning} and therefore it is an unlikely explanation for this outlier.\par
Having provided the ANITA-I flight simulation with the most realistic treatment of radio signals, we investigate the impact of the corrections on the electric field values from ZHAireS simulations. Not including the correction due to the reflection on the curved ice surface results in a significant increase of sensitivity for events near the horizon; therefore, the total number of expected events increased to 21 using the Pierre Auger Observatory spectrum and 25 for the Telescope Array spectrum. Assuming a completely smooth surface results in tension between the simulated and observed distributions of the spectral slope: $P_{KS}$ = 0.03 using the Pierre Auger Observatory energy spectrum and $P_{KS}$ = 0.01 when using the Telescope Array energy spectrum. These results underline the importance of having a correct treatment of the reflection. Some studies of the reflection properties of the Antarctic ice in the ANITA-I frequency band have been performed using the direct and reflected emission from the Sun \cite{ANITA_REFL}. These studies indicate that the reflection coefficient roughly follows the expectation for Fresnel coefficients at low incident angles, but smaller values than predicted by Fresnel coefficients are observed at larger angles of incidence. This is qualitatively what is expected for both the defocusing due to surface curvature and the loss of coherence due to surface roughness. However, to exactly calculate the reflective properties of the ice the location dependent surface model needs improvement.  Additional refinement of the models could include local curvature, local slope and local variation in refractivity of the surface. However, to generate these models additional measurements will be necessary. During the third ANITA flight (2014-2015) a companion payload equipped with a pulser flew for exactly this purpose. This payload transmitted pulses that can be observed directly and after reflection from the surface. \par
A small fraction of Antarctic surface has hills and mountains that might influence the reflection of the radio signal. These intermediate length scale effects are not taken into account in the analysis in this paper. The pulser campaign of the third ANITA flight might shed light on the influence of surface effects on these length scales.\par

There have only been a few comparisons between observed and simulated radio signals from air showers and mainly at lower frequencies than the frequency band of ANITA. Quantitatively, good agreement is achieved with recent observations at the Low Frequency Array (LOFAR) \cite{ref:LOFAR190,ref:LOFAR30}, however an absolute calibration is not provided. Comparisons including an absolute calibration vary between $\sim$10\,\% \cite{ref:AugerICRC} and $\sim$16\,\% \cite{ref:LOPES,ref:LOPESCOR} deviations between simulations and observations. Recently, the CROME experiment observed several air showers using gigahertz receivers \cite{ref:CROME}. The distribution of shower impact locations was in agreement with expectations from simulations. Which indicates that current simulation models are capable of describing radio emission over a large frequency range.\par
The  agreement that was found between number of predicted events from simulation and the observed number of events is encouraging for the validity of the absolute scale of the radio signal from the simulations.  However,  in the near future the radio emission codes will be absolutely calibrated by the antenna arrays incorporated in one of the existing cosmic-ray observatories (LOFAR \cite{ref:LOFAR30}, AERA \cite{ref:AugePol}, TunkaRex \cite{TunkaRex}, CODALEMA \cite{ref:CodGeo}, and by dedicated particle-beam experiments such as T-510 at SLAC \cite{ref:T510}.\par
The energy scale of the analysis presented here is significantly lower than the first results published on this data set \cite{ANITA_CR}. The main reason for this discrepancy is that at the time of that publication there was no radio emission model known that could produce coherent radiation in the ANITA frequency band. Therefore all observables were fitted together in a Bayesian approach with a simple parameterization of the radio beam. The input parameters were based on observations made at lower frequencies. Therefore, the fit to the data preferred a wider beam with significantly less power at higher frequencies than what the models predict now resulting in higher energy estimates and larger exposure at high energy. However, the recent incorporation of the Cherenkov-like effects in the emission models significantly altered the lateral and frequency dependencies of the radio beam. We have shown that we now can independently estimate the energy of the cosmic rays and get reasonable agreement between the distributions from observations and flight simulations. Therefore,  a parameterized fit is no longer necessary and both cosmic-ray energy and the detector exposure can be obtained directly from simulations. 

\section{Conclusion and Outlook}
In this paper, we present a new method of energy reconstruction for cosmic rays using observations in the frequency range of 300 and 1000 MHz and applied this to observations made during the ANITA-I flight. The mean energy of cosmic rays  observed by the first ANITA flight is  $2.9 \pm 0.4 (\sigma_i) \pm 0.8 (\sigma_s)\times10^{18} $\,eV.\par 
The analysis in this paper shows that by using antennas sensitive between 300-1000\,MHz there is a straightforward method to estimate the energy of the cosmic-ray particle. Since this method of energy estimation is independent of energy estimation from other techniques it might become useful in the cross calibrating of the different techniques when implemented in existing cosmic-ray observatories.\par
We simulated the full ANITA flight to estimate the sensitivity to cosmic rays that produce radio signals that are reflected from the ice. From this simulation, we find that the expected number of events is in agreement with the observed number of events. We estimated the total exposure of the flight and estimated, for the first time, the cosmic-ray flux based on radio observations only. This cosmic-ray flux estimate is in agreement with observations made at the Pierre Auger Observatory and Telescope Array.\par
The flight simulation is also used to compare the distributions of simulated events that trigger the ANITA instrument to the distributions obtained from the measurements. We found reasonable agreement between simulations and measurements in the distribution of the incident angle, the spectral slope, and the amplitude. This provides us with some confidence that the simulations are consistent and allows us to study the importance and necessity of the assumptions made in the flight simulation. However, the limited size of our data set prevents us to go into very deep details and therefore we provide a single flux point rather than an energy spectrum. In the 2014-2015 Austral summer the third ANITA flight flew with a significantly higher sensitivity for cosmic-ray signals than the previous flights. Therefore, a significant increase of the number of cosmic-ray observations is expected. This data set will allow for an energy spectrum and study more of the details of the reflections. \par
The exposure of the ANITA-I flight is orders of magnitude too small to significantly contribute to the collection of cosmic rays with energy above the GZK limit ($\sim5 \times 10^{19}$ eV). However, it was shown that with a measurement at a single location with respect to the shower axis we were able to measure the cosmic-ray energy with a reasonable accuracy. Therefore, with an array of antennas on the ground, broadband observations might be used to constrain the energy and other parameters, such as the longitudinal air shower development \cite{ref:KrijnXMAX}, with only a few antennas per air shower. It should be noted that the accuracy to determine air shower properties is in the case of ANITA largely set by uncertainties on the reflection properties of the surface, these can of course be discarded in the case of a ground array of antennas. In addition, a ground array will point up to the sky and therefore the antenna temperature is expected to drop significantly with respect to the downward pointing antennas of ANITA. At lower frequencies (50\,MHz) the antenna temperature is more than two orders magnitude higher than in the ANITA frequency range due to the galactic background. The straightforward way of deriving shower parameters with only a few broadband antennas in combination with the low antenna temperature are beneficial characteristics to be utilized in a cosmic-ray ground array. However, since the footprint on the ground is rather small, it is unlikely that this technique is scalable to a large array for detecting UHECRs at a low cost. However, it could be either incorporated in an existing cosmic-ray observatory to provide a cross-calibration technique for the energy estimation of air showers or as an array making accurate measurements of air showers around $10^{16}$ - 10$^{18}$\,eV.\par

\section*{Acknowledgments}
We would like to thank Marianne Ludwig from Karlsruhe Institute of Technology for helpful discussions and in the early days of this work. We are grateful to NASA, the U.S. National Science Foundation, the U.S. Department of Energy, and the Columbia Scientific Balloon Facility for their generous support of these efforts. We would like to extend our thanks to the  2006-2007 on-ice LDB and McMurdo crews for their support. Part of the research was carried out at the Jet Propulsion Laboratory, California Institute of Technology, under a contract with the National Aeronautics and Space Administration. J. A-M, W.R.C., D.G-F and E.Z. thank Ministerio de Econom\'\i a (FPA2012-39489), Consolider-Ingenio 2010 CPAN Programme (CSD2007-00042),
Xunta de Galicia (GRC2013-024), Feder Fundsand Marie Curie-IRSES/ EPLANET (European Particle physics Latin American NETwork), $7^{\rm th}$ Framework Program (PIRSES-
2009-GA-246806).

\clearpage

\section*{Appendix A: Frequency dependency of surface roughness reflection coefficient}

To address the effect of surface roughness we followed the method as described in \cite{ref:SWORD}. As an input we take a parameterization of the surface roughness, as a function of length scales, that is fitted to observations from \cite{RADARSAT} and measurements taken at the Antarctic Taylor Dome Station.\par
\begin{figure}
\begin{center}
{
\includegraphics[width=0.7\textwidth]{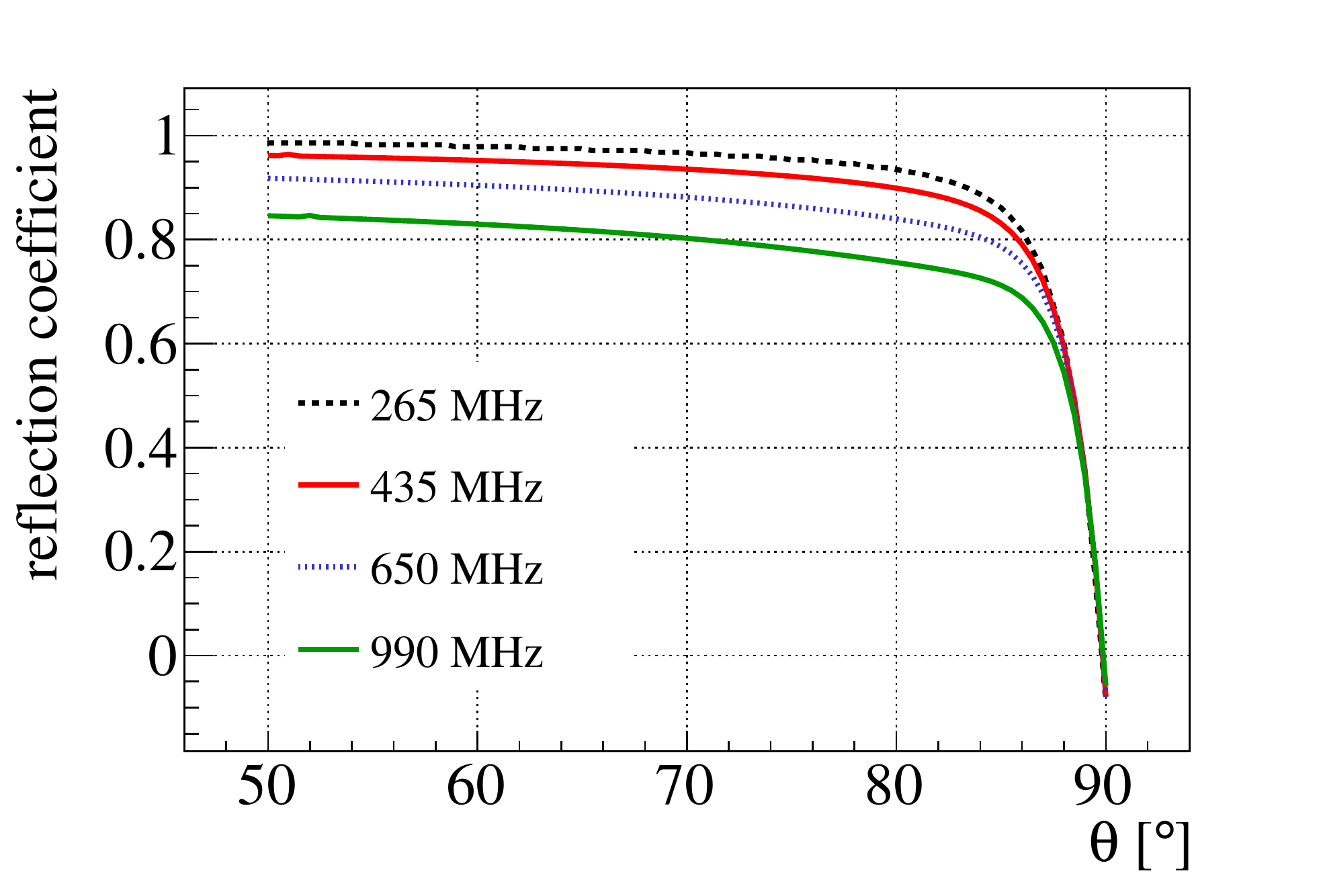} 
}
\caption{The amplitude multiplication factors due to surface roughness. They are shown for four frequencies within the ANITA frequency band as a function of incident angle on the ice.}
\label{fig:rough}
\end{center}
\end{figure}
The calculation of the frequency dependent roughness effect is very CPU time consuming. Therefore, it has been calculated at a few specific frequencies and over a range of incident angles. Afterwards, the results are parametrized with smooth functions to intermediate distance values. The multiplication factors as a function of incident angle are shown in Figure \ref{fig:rough} at four different frequencies. We perform a linear interpolation between the results for the different frequencies to obtain a frequency dependent multiplication factor.
 
\section*{Appendix B: Comparing ZHAireS to CoREAS}
The ZHAireS simulation package used in this paper is the only one that can estimate the electric field at location of the payload after reflection 
on the ice cap \cite{ref:ZHSreflex}. However, we made a comparison to the CoREAS package \cite{ref:CoREAS} to estimate the discrepancy 
between the two. This forced us to perform calculations on locations on the ground only in order to compare them.\par
The implementation of the \emph{end-point} algorithm for the calculation of the electric field in CoREAS and the corresponding \emph{ZHS} 
algorithm in ZHAireS result in identical amounts of radiation when applied to identical charge distributions. However, since the two algorithms 
are implemented in different air shower simulation packages, differences in the calculated radiation may occur. To minimize these differences 
we modified the CoREAS code in order to use the same model of refractive index as in ZHAireS which was tuned to Antarctic atmospheric data. 
We also selected air showers from both simulations that have $X_{\text{max}}$ that deviates less than 5 g\,cm$^{-2}$ from a fixed value. \par
\begin{figure}
\begin{center}
{
\includegraphics[width=0.95\textwidth]{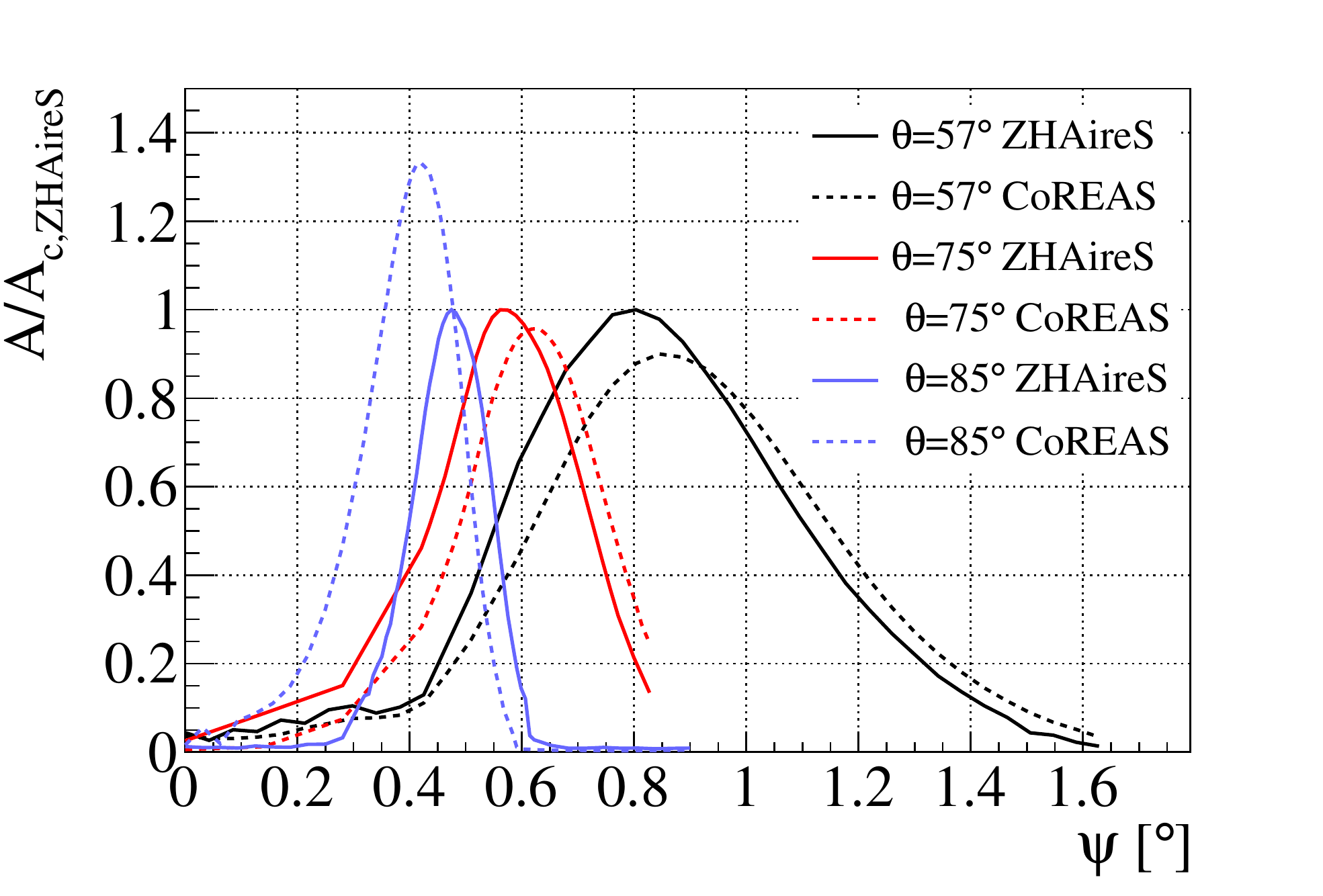} 
}
\caption{Comparison of $A$ values from CoREAS and ZHAireS on the ground for a 57$^\circ$ zenith angle shower with cosmic-ray energy of 10$^{17}$ eV.  We normalized the $A$ values to that obtained from ZHAireS on the Cherenkov cone $A_{c,\text{ZHAireS}}$. The lines correspond 
to an average value of $A$ obtained from a set of 8 to 25 simulations (depending on the geometry).}
\label{fig:CoREASZHS}
\end{center}
\end{figure}
In Figure \ref{fig:CoREASZHS}, we compare the distributions of $A$ as a function of $\psi$ (off-axis angle) on the ground 
as obtained with ZHAireS and CoREAS for three different shower zenith angles. 
Overall the distributions in amplitude look rather similar for the two packages. The 10\% and 5\% deviations observed at the two smallest 
angles might be in part due to different cuts on particle energy while simulating the air showers with both packages. The 30\% deviation 
at the largest zenith angle may have several small causes that get emphasized at extreme zenith angles.\par
We adopt the observed differences between the two simulations packages as an uncertainty on the field strength 
due to the simulation of the radio signal.


\begin{thebibliography}{99}

\bibitem{ref:AugerSD} The Pierre Auger Collaboration, \emph{The surface detector system of the Pierre Auger Observatory}, Nuclear Instruments and Methods in Physics Research Section A: Accelerators, Spectrometers, Detectors and Associated Equipment, Volume 586, Issue 3, 1 March 2008, Pages 409-420, ISSN 0168-9002

\bibitem{ref:AugerFD} The Pierre Auger Collaboration, \emph{The fluorescence detector of the Pierre Auger Observatory}, Nuclear Instruments and Methods in Physics Research Section A: Accelerators, Spectrometers, Detectors and Associated Equipment, Volume 620, Issues 2-3, 11-21 August 2010, Pages 227-251, ISSN 0168-9002 

\bibitem{ref:TASD} T. Abu-Zayyad \emph{et al} ,\emph{The surface detector array of the Telescope Array experiment}, Nuclear Instruments and Methods in Physics Research Section A: Accelerators, Spectrometers, Detectors and Associated Equipment, Volume 689, 11 October 2012, Pages 87-97, ISSN 0168-9002,

\bibitem{ref:TAFD} H. Tokuno, Y. Tameda \emph{et al},\emph{New air fluorescence detectors employed in the Telescope Array experiment}, Nuclear Instruments and Methods in Physics Research Section A: Accelerators, Spectrometers, Detectors and Associated Equipment, Volume 676, 1 June 2012, Pages 54-65, ISSN 0168-9002

\bibitem{ref:TA} T. Abu-Zayyad et al (Telescope Array Collaboration), \emph{Energy spectrum of ultra-high energy cosmic rays observed with the Telescope Array using a hybrid technique},Astroparticle. Physics.  61, 93 (2015).

\bibitem{ANITA_CR}S. Hoover \emph{et al},  \emph{Observation of Ultra-high-Energy Cosmic Rays with the ANITA Balloon-Borne Radio Interferometer}, PRL {\bf{105}}, (2010)

\bibitem{ref:CodGeo}
D. Ardouin et al,\emph{Geomagnetic origin of the radio emission from cosmic ray induced air showers observed by CODALEMA}, Astroparticle Physics, Volume 31, Issue 3, April 2009, Pages 192-200

\bibitem{ref:AugePol}A. Aab et al. (Pierre Auger Collaboration),\emph{Probing the radio emission from air showers with polarization measurements}
Phys. Rev. D 89, 052002, Published 14 March 2014

\bibitem{ref:AnitaExp} P.W. Gorham \emph{et al,} \emph{The Antarctic Impulsive Transient Antenna ultra-high energy neutrino detector: Design, performance, and sensitivity for the 2006-2007 balloon flight}, Astroparticle Physics, Volume 32, Issue 1, August 2009, Pages 10-41, ISSN 0927-6505,



\bibitem{ref:AndresInfer} A. Romero-Wolf, S. Hoover, A.G. Vieregg, P.W. Gorham \emph{et al},\emph{An interferometric analysis method for radio impulses from ultra-high energy particle showers}, Astroparticle Physics, Volume 60, January 2015, Pages 72-85, ISSN 0927-6505


\bibitem{ref:ZHAireSCher}J. Alvarez-Mu\~niz, W. R. Carvalho Jr., A. Romero-Wolf, M. Tueros, E. Zas, \emph{Coherent Radiation from Extensive Air Showers in the Ultra-High Frequency Band }, Phys. Rev. D86, 123007, 2012 

\bibitem{ref:KrijnCher}K. D. de Vries, A. M. van den Berg, O. Scholten, and K. Werner, \emph{Coherent Cherenkov Radiation from Cosmic-Ray-Induced Air Showers}, Phys. Rev. Lett. 107, 061101 - Published 2 August 2011

\bibitem{ref:CoREAS} T. Huege, M. Ludwig, C. W. James, \emph{Simulating radio emission from air showers with CoREAS}, (2013)	\url{arXiv:1301.2132}

\bibitem{ref:SELFAS} Benoit Revenu, Vincent Marin, \emph{Coherent radio emission from the cosmic ray air shower sudden death},Proceedings of  33rd International Cosmic Ray Conference, Rio De Janeiro2013, \url{arXiv:1307.5673} 

\bibitem{ref:ZHSreflex} J. Alvarez-Mu\~niz, W. R. Carvalho Jr.,
D. Garc{\'\i}a-Fern\'andez, H. Schoorlemmer, E. Zas.
\emph{Simulations of reflected radio signals from cosmic ray
induced air showers}, Astroparticle Physics 66 (2015) 31-38.

\bibitem{ref:ZHAireS} J. Alvarez-Mu\~niz, W. R. Carvalho Jr., E. Zas,\emph{ Monte Carlo simulations of radio pulses in atmospheric showers using ZHAireS}, Astroparticle Physics, Volume 35, Issue 6, January 2012, Pages 325-341

\bibitem{ref:Kon} K. Belov and ANITA Collaboration,\emph{Towards determining the energy of the UHECRs observed by the ANITA detector},AIP Conf. Proc. 1535, 209 (2013),  Erlangen, Germany

\bibitem{ref:ZAS} J. Alvarez-Mu\~niz, R. A. V\'azquez and E. Zas, 
\emph{Characterization of neutrino signals with radio pulses in dense media
through the Landau-Pomeranchuk-Migdal effect}
Phys. Rev. D, {\bf 61}, 023001 (1999) 

\bibitem{ref:ZHS2com}J. Alvarez-Mu\~niz, W. R. Carvalho Jr., H. Schoorlemmer, E. Zas,\emph{Radio pulses from ultra-high energy atmospheric showers as the superposition of Askaryan and geomagnetic mechanisms}, Astroparticle Physics, Volume 59, July-August 2014, Pages 29-38, ISSN 0927-6505

\bibitem{andres} A. Romero-Wolf, S. Vance, F. Maiwald,
E. Heggy, P. Ries, K. Liewer. \emph{A passive probe for
subsurface oceans and liquid water in Jupiter’s icy moons},
Icarus 248, 463-477 (2015).

\bibitem{ref:SWORD} A. Romero-Wolf, P. Gorham, K. Liewer, J. Booth, R. Duren, \emph{Concept and Analysis of a Satellite for Space-based Radio Detection of Ultra-high Energy Cosmic Rays}, 	\url{arXiv:1302.1263} [astro-ph.IM], (2013)


\bibitem{ref:radarscatter} M.K. Shepard and B.A. Campbell, \emph{
Radar Scattering from a Self-Affine Fractal Surface: Near-Nadir Regime}, Icarus,{\bf{141}}, 156-171 (1999)


\bibitem{RADARSAT}H. Liu, K. Jezek, B.Li,and Z. Zhao.2001. Radarsat Antarctic Mapping Project digital elevation model version
2. Boulder, CO: National Snow and Ice Data Center. Digital media.


\bibitem{ref:AugerXmax} A. Aab et al. (Pierre Auger Collaboration),\emph{Depth of maximum of air-shower profiles at the Pierre Auger Observatory. I. Measurements at energies above 10$^{17.8}$\,eV},Phys. Rev. D 90, 122005, Published 31 December 2014

\bibitem{ref:priv} P. Motloch, N. Hollon, P. Privitera, \emph{On the prospects of Ultra-High Energy Cosmic Rays detection by high altitude antennas}, Astroparticle Physics, Volume 54, February 2014, Pages 40-43, ISSN 0927-6505

\bibitem{ref:AugerICRC} The Pierre Auger Collaboration, \emph{The Pierre Auger Observatory: Contributions to the 33rd International Cosmic Ray Conference (ICRC 2013)}, (2013)	\url{arXiv:1307.5059}

\bibitem{ref:ThunMand} N. Mandolesi, G. Morigi, and G. Palumbo, \emph{Radio pulses from extensive air showers during thunderstorms, the atmospheric electric field as a possible cause}, Journal of Atmospheric and Terrestrial Physics 36 (1974), no. 8, 1431 ? 1435. 

\bibitem{ref:ThunLOPES} S. Buitink et al., \emph{Amplified radio emission from cosmic ray air showers in thunderstorms}, A\&A 467 (2007) no. 2, 385?394

\bibitem{ref:ThunLOFAR}P. Schellart et al., \emph{Probing Atmospheric Electric Fields in Thunderstorms through Radio Emission from Cosmic-Ray-Induced Air Showers},Phys. Rev. Lett. 114, 165001,24 April 2015

\bibitem{ref:NASAlighning} Cecil, D.J., D. Buechler, and R. Blakeslee. 2014. LIS/OTD Gridded Lightning Climatology Data Sets. Data set available online \url{ftp://ghrc.nsstc.nasa.gov/pub/lis/climatology} from the NASA Global Hydrology Resource Center DAAC, Huntsville, Alabama, U.S.A. doi: http:/dx.doi.org/10.5067/LIS/LIS-OTD/DATA311

\bibitem{ANITA_REFL} Besson, D. Z., et al. (2015), \emph{Antarctic radio frequency albedo and implications for cosmic ray reconstruction}, Radio Sci., 50, 1-17, doi:10.1002/2013RS005315.


\bibitem{ref:LOFAR190}A. Nelles, S. Buitink, H. Falcke, J. R. H\"orandel, T. Huege, P. Schellart,\emph{A parameterization for the radio emission of air showers as predicted by CoREAS simulations and applied to LOFAR measurements}, Astroparticle Physics, Volume 60, January 2015, Pages 13-24, ISSN 0927-650

\bibitem{ref:LOFAR30} A. Nelles \emph{et al}, \emph{Measuring a Cherenkov ring in the radio emission from air showers at 110-190 MHz with LOFAR}, Astroparticle Physics, Volume 65, May 2015, Pages 11-21, ISSN 0927-6505

\bibitem{ref:LOPES} W.D. Apel \emph{et al}, \emph{Comparing LOPES measurements of air-shower radio emission with REAS 3.11 and CoREAS simulations}, Astroparticle Physics, Volumes 50-52, December 2013, Pages 76-91, ISSN 0927-6505, 

\bibitem{ref:LOPESCOR} W.D. Apel \emph{et al}, \emph{Improved absolute calibration of LOPES measurements and its impact on the comparison with REAS 3.11 and CoREAS simulations}, arXiv:1507.07389

\bibitem{ref:CROME}R. \u{S}m\'ida, \emph{et al}, \emph{First Experimental Characterization of Microwave Emission from Cosmic Ray Air Showers}, Phys. Rev. Lett. 113, 221101, Nov. 2014

\bibitem{TunkaRex} P.A. Bezyazeekov et al. - Tunka-Rex Collaboration, \emph{Measurement of cosmic-ray air showers with the Tunka Radio Extension (Tunka-Rex)} Nucl. Instr. Meth. A 802 (2015), 89-96

\bibitem{ref:T510} K. Belov, K. Mulrey, A. Romero-Wolf, S. A. Wissel, A. Zilles, \emph{et al} , \emph{Accelerator measurements of magnetically-induced radio emission from particle cascades with applications to cosmic-ray air showers}, 	arXiv:1507.07296, July 2015


\bibitem{ref:KrijnXMAX}K. D. de Vries, O. Scholten, K. Werner, \emph{The air shower maximum probed by Cherenkov effects from radio emission}, Astroparticle Physics, Volume 45, May 2013, Pages 23-27
\end{thebibliography}
\end{document}